\documentclass[11pt]{article}
\usepackage{amssymb,amsmath,mathrsfs,enumitem, amsthm,xspace,graphicx,authblk}%,showkeys}

\theoremstyle{definition}
\newtheorem{definition}{definition}[section]
\theoremstyle{plain}
\newtheorem{theorem}[definition]{Theorem}

\theoremstyle{remark}

\newcommand{\bR}{\mathbb{R}}
\newcommand{\bS}{\mathbb{S}}
\newcommand\nth{\textsuperscript{th}\xspace}
\newcommand\bq{{\boldsymbol q}}
\DeclareMathOperator{\jSp}{jSp}
\newcommand{\vnorm}[1]{{\pmb{|}}\,#1\,{\pmb{|}}}
\numberwithin{equation}{section}

\author[1]{Dorothea Bahns}
\author[2]{Sergio Doplicher}
\author[3]{Gerardo Morsella}
\author[4]{Gherardo Piacitelli}
\affil[1]{\small Mathematisches Institut and
 Courant Research Centre ``Higher Order Structures in Mathematics", Universit\"at G\"ottingen\\
 
 Bunsenstr. 3-5, D-37073 G\"ottingen (Germany), e-mail: bahns@uni-math.gwdg.de}
\affil[2]{\small Dipartimento di Matematica, Universit\`a di Roma ``La Sapienza''\\

p.le Aldo Moro, 5, I-00185 Roma (Italy), e-mail: dopliche@mat.uniroma1.it}
\affil[3]{\small Dipartimento di Matematica, Universit\`a di Roma Tor Vergata\\

v.le della Ricerca Scientifica, 1, I-00133 Roma (Italy)\\

e-mail: morsella@mat.uniroma2.it}
\affil[4]{\small SISSA, Via Bonomea 265,  I-34136 Trieste (Italy), e-mail: piacitel@sissa.it}

\title{QUANTUM SPACETIME AND ALGEBRAIC QUANTUM FIELD THEORY}

\begin{document}
\maketitle

\begin{abstract} 
We review the investigations on the quantum structure of spactime, to be found at the Planck scale if one takes into account the operational limitations to localization of events which result from the concurrence of Quantum Mechanics and General Relativity. We also discuss the different approaches to (perturbative) Quantum Field Theory on Quantum Spacetime, and some of the possible cosmological consequences.
\end{abstract}

\tableofcontents

\section{Quantum nature of spacetime at the Planck scale: why and how} \label{sec:why}

According to Classical General Relativity, at large scales spacetime is a pseudo Riemanniann manifold locally modelled
on Minkowski space. But the concurrence with the principles of Quantum Mechanics renders this picture untenable {\itshape in the small.}

Those theories are often reported as hardly reconcilable, but they do meet
at least in a single {\itshape partial} principle, the 
{\itshape Principle of Gravitational Stability against localisation of events} 
formulated in
\cite{dopl1, dopl2}:
\begin{quote}
{\itshape The gravitational field generated by the concentration of energy
required by the Heisenberg Uncertainty Principle to localise an event in
spacetime should not be so strong to hide the event itself to any distant
observer - distant compared to the Planck scale.}
\end{quote}

The effect  of this principle is best seen considering first the effect of an observation which 
locates an event, say, in a spherically symmetric way around the origin in 
space with accuracy \(a\); according to Heisenberg principle an uncontrollable 
energy \(E\) of order \(1/a\) has to be transferred, which will generate a 
gravitational field with Schwarzschild radius \(R \simeq E\) (in universal units where \( \hbar = c =  
G  = 1\)). Hence we {\itshape must} have that \(a \gtrsim R \simeq  1/a\); so that  \(a
\gtrsim 1\), i.e. in CGS units

\begin{equation}
\label{dopleq11}
a \gtrsim \lambda_P \simeq 1.6 \cdot 10^{-33} cm.
\end{equation}

This folklore argument is certainly very old, but its elaborations in two significant directions are surprisingly recent.

First, if we consider generic uncertainties, the argument above suggests that they ought to be limited by uncertainty relations.

Indeed, if we measure one of the space coordinates of our event with 
great precision \(a\), but allow large uncertainties \(L\) in the knowledge 
of the other coordinates, the energy \(1/a\) may spread over a thin disk of 
radius \(L\) and thus generate a gravitational potential that would vanish 
everywhere as \(L \rightarrow \infty\) (provided \(a\), as small as we like but non zero, remains constant).

This is shown by  trivial computation of the Newtonian potential generated by the corresponding {\itshape mass} distribution; whenever such a potential is nearly vanishing, nobody would expect large General Relativistic or Quantum Gravitational corrections; so we can rely on that estimate.

An equally elementary computation would show that the same conclusion holds if {\it two} space coordinates are measured with small but fixed precision \(a\) and the third one with an uncertainty \(L\), and \(L \rightarrow \infty\). 

Second, if we consider the energy content of a generic quantum state where the location measurement is performed, the bounds on the uncertainties should depend also upon that energy content \cite{dopl7, dopl13} .

To see this point, just suppose that our background state describes the spherically symmetric distribution of the total
energy \(E\) within a sphere of radius \(R\), with \(E  <  R\). If we localise,
in a spherically symmetric way, an event at the origin with space accuracy
\(a\), due to the Heisenberg Principle the total energy will be of the order
\(1/a  +  E\). We must then have
\[
\frac{1}{a}  +  E   <   R,
\]
otherwise our event will be hidden to an observer located far away, out of
the sphere of radius \(R\) around the origin.
Thus, if \(R  -  E\) is much smaller than \(1\), the ``minimal distance'' will
be much larger than \(1\).
But if \(a\) is anyway larger than \(R\) the condition implies rather
\[
\frac{1}{a}  +  E   <   a.
\]
Thus, if \(R  -  E\) is very small compared to \(1\) and \(R\) is much larger
than \(1\), \(a\) cannot be essentially smaller than \(R\).

Now the causal relations between events should also break down at scales which are so small that events cannot be localised that sharply; hence we have to expect that scale to express the range of propagation of acausal effects. 

This naive picture suggests that, due to the principle of
Gravitational Stability,  initially all points of the Universe should have been
causally connected.

Thus we can expect that Quantum Spacetime (QST) {\itshape solves} the horizon problem (cf.~\cite{dopl7} for hints in that direction, \cite{dopl13} or Section~\ref{subsec:horizon} below for an indication that a Quantum Spacetime with a constant Planck length should generate dynamically a range of propagation of acausal effects which solves the horizon problem).
 
We come back to the general discussion. If we aim at a merge of Quantum Mechanics and General Relativity we should reason in terms of concepts which are physically legitimate from the general relativistic point of view as well. One might doubt from the start about concepts like {\itshape local energy and coordinates} to which the Heisenberg Principle refers.

Concerning the use of coordinates, one should better talk of measurements conditioned to the measurement of a finite number of auxiliary local quantities; in some appropriate limit, in Minkowski space, that auxiliary measurement should become the specification of a frame. Thus the use of coordinates should be legitimate at a semiclassical level.

Another important reason to work with coordinates is that we are interested in the tangent space at a point equipped with normal coordinates, describing a free falling system in Einstein's lift. Or a system in a constant gravitational field; for the outside distribution of matter on the large scale, such as the structure of the Virgo supercluster of galaxies to which we belong, ought to have no influence on a high energy collision in the CERN collider; even if we were so clever to detect (quantum) effects of the gravitational forces {\itshape between} the colliding particles.

Thus in a first stage it is legitimate, and physically reasonable, to study the small scale structure of Minkowski space. The spacetime symmetries of our space ought to be described by the {\itshape classical} Poincar\'e group: for the {\itshape global motions} of our space should look the same in the large as they do in the small, and, in the large, they should be precisely the {\itshape classical symmetries.}  

One other remark in order here concerns the very nature of the coordinates.  In the Quantum Mechanics of systems with finitely many degrees of freedom, they are {\itshape observables} describing the particle positions.

In Quantum Field Theory, the observables are {\itshape local} quantities associated each with a finite region in spacetime. They can never describe {\itshape exactly} a property of one particle or \(n\) - particle states, which are global (asymptotic) constructs. If that region reduces to a point, we find only the multiples of the identity. We ought to consider open regions. We might consider such a region as a neighbourhood of a spacetime point, defining it with some uncertainty, and the measurement of associated local quantities as leading to information on that location.

Thus Spacetime appears as a space of parameters, which, in absence of gravitational forces, can be specified with arbitrarily high (but finite!) precision, with higher and higher energy cost for higher and higher precision. The consideration of the gravitational effects of that energy cost will cause, as we will see, that space of parameters to become {\itshape noncommutative}.

The semiclassical level of a first analysis justifies also the use of concepts like energy; but a more careful analysis shows, as briefly mentioned here in the sequel, that  in essence the conclusions remain true without any reference to the concept of energy.

At a semiclassical level, the main consequence of the Principle stated above is the validity of  {\itshape Spacetime
Uncertainty Relations}; furthermore, they have been shown to be implemented by
Commutation Relations between coordinates, thus turning Spacetime into
{\itshape Quantum Spacetime} \cite{dopl1,dopl2}. 

The word ``Quantum'' is very
appropriate here, to stress that noncommutativity does not enter just as a
formal generalisation, but is strongly suggested by a compelling
{\itshape physical} reason, unlike the very first discussions of possible
noncommutativity of coordinates in the pre-renormalisation era, by
Heisenberg, Snyder and Yang, where noncommutativity was regarded as a
curious, in itself physically doubtful, possible {\itshape regularisation device}, without any reference to General Relativity and Gravitational forces; the qualitative fact that the quantum structure of gravitational forces ought to have consequences on the nature of spacetime in the small was anticipated by P.M.Bronstein \cite{dopl14}, where, however, the focus was on the extension of the Bohr-Rosenfeld argument to the Christoffel symbols, and on the proposal of a Quantum Theory of {\itshape linearised} Gravity, without any mention of spacetime uncertainty relations. 

The analysis based on the  Principle of Gravitational Stability against localisation of events leads to the following conclusions:
\begin{enumerate}[label=\roman*)]
\item There is no a priori lower limit on the precision in the
measurement of any {\itshape single} coordinate (it is worthwhile to stress once
more that the apparently opposite
conclusions, still often reported in the literature in connection with the ACV variant of the Heisenberg principle \cite{ACV}, are drawn under the
$implicit$ assumption that {\itshape all} the space coordinates of the event are
simultaneously sharply measured).

Every alerted reader will note that nobody knows an operational prescription to measure, say, {\itshape only one spacetime coordinate} of the location of an event with a terrific (ultra Planckian) precision. But of course we cannot say that such a measurement is {\itshape impossible} just because we are not capable of inventing a device; we could say that only if we could show that it is {\itshape forbidden} by the presently known physical principles. Which at present does {\itshape not} seem to be the case.

\item The uncertainties  \(\Delta q_\mu\) in the measurement of the coordinates of an event in Minkowski space should be at least bounded by the following Spacetime Uncertainty Relations:
\begin{subequations}\label{eq:stur}
\begin{gather}\label{eq:stur_electric}
\Delta q_0 \cdot \sum \limits_{j = 1}^3 \Delta q_j \gtrsim 1 ;\\ 
\label{eq:stur_magnetic}
\sum\limits_{1 \leq j < k \leq 3 } \Delta q_j  \Delta q_k \gtrsim 1 .
\end{gather}
\end{subequations}

Thus points become fuzzy and {\itshape locality looses any precise meaning}.
We believe it should be replaced at the Planck scale by an equally sharp and 
compelling principle, which reduces to locality at larger distances. Such a principle is 
nowadays totally unknown, and unaccessible by operational reasoning. 
\end{enumerate}

Some comments on the derivation of these relations are in order. In the analysis of 1994--95, they were justified {\itshape in special cases} by their consistency with the {\itshape exact} solutions of Einstein Equations (EE), as Schwarzschild and Kerr's solutions. But in general they were derived using the {\itshape linearised} approximation to EE.

Furthermore the concept of energy was central: in a {\itshape semiclassical} approach, the expectation value in a state describing an {\itshape ansatz} for the outcome of a localisation experiment (a coherent state in a free field theory) of the {\itshape energy-momentum tensor} for that field, was used as a source for the linearised EE.   

Then, the requirement of non-formation of trapped surfaces hiding the observed event  was formulated as the condition of non negativity of the time-time component of the metric tensor.  The relations above follow as a weaker simplified necessary condition.

Both the use of the linearised approximation and of the notion of energy are doubtful.

But in recent works  \cite{dopl11,dopl12}
 Tomassini and Viaggiu have shown that (a stronger form of) the above relations do follow from an {\itshape exact} treatment, if one adopts the {\itshape Hoop Conjecture}, which limits the energy content of a space volume in terms of the area of the boundary, as a condition for the non-formation of bounded trapped surfaces. Moreover, their analysis applies to a curved background as well.

The treatment is again semiclassical, and involves the notion of energy, but the conflict about the use of the linearised approximations to derive bounds, and imposing those bounds in situations close to singularities, disappears.

Eventually, in \cite{dopl13} the special case of spherically symmetric experiments, with all spacetime uncertainties taking the same value, was treated with use of  the {\itshape exact} semiclassical EE, without any reference to the energy observables.  The state describing the outcome of the localisation experiment was taken {\itshape not} as a strictly localised state, but as the state, with weaker localisation properties, obtained acting on the vacuum state with the field operators themselves, smeared with test functions having the appropriate symmetry, in a theory of a single scalar massless field coupled semiclassically to gravity. The solution of the Raychaudhuri equation yields to the universal lower bound for the common value of the uncertainties, of the order of Planck length (see also Section~\ref{subsec:spherical} below for more details). We stress that this result gives a possibly weaker condition than the condition which could be derived by a choice of better localised {\itshape ans\"atze} for the probe state.

We can conclude that the above Spacetime Uncertainty Relations are reasonably well grounded for Minkowski space; they are to be expected to hold in similar variant in curved spacetimes, by the Tomassini-Viaggiu argument; a basic consequence of those relations, when implemented by the Quantum Conditions we will now discuss, namely that the Planck scale is a universal minimal length, is well grounded on the basis of the most general assumptions, in the spherically symmetric case.

The Spacetime Uncertainty Relations strongly suggest that spacetime has a
{\itshape Quantum Structure} at small scales, expressed, in generic units, by
\begin{equation}
\label{dopleq13}
       [q_\mu  ,q_\nu  ]   =   i \lambda_P^2   Q_{\mu \nu},
\end{equation}
where \(Q\) has to be chosen not as a random toy mathematical model, but in
such a way that (\ref{eq:stur}) follows from (\ref{dopleq13}).

To achieve this in the simplest way, it suffices to select the model where 
the \(Q_{\mu \nu}\) are {\itshape central}, and impose the ``Quantum Conditions'' on the 
two invariants
 
\begin{equation}
\label{dopleq14}
Q_{\mu \nu} Q^{\mu \nu};
\end{equation}
\begin{align}
\left[q_0  ,\dots,q_3 \right]  &\equiv  \det \left(
\begin{array}{ccc}
q_0 & \cdots  & q_3 \nonumber\\
\vdots  & \ddots  & \vdots  \\
q_0 & \cdots  & q_3
\end{array}
\right)\\
&\equiv  \varepsilon^{\mu \nu \lambda \rho} q_\mu q_\nu q_\lambda q_\rho =\nonumber\\
&= - (1/2) Q_{\mu \nu}  (*Q)^{\mu \nu};\label{dopleq15} 
\end{align}
whereby the first one must be zero and the square of the half of the second is \(I\) (in 
Planck units; we must take the square since it is a pseudoscalar and not a 
scalar).

One obtains in this way \cite{dopl1,dopl2}  a model of Quantum Spacetime 
which implements 
exactly our Spacetime Uncertainty Relations and is fully Poincar\'e 
covariant. 

As anticipated, here the {\itshape classical} Poincar\'e group acts as symmetries; translations, in particular, act adding to each \(q_\mu\) a real multiple of the identity.

Thus ``coordinates'' and ``translation parameters'', classically described by the same objects, hear split into different entities;  but this happens already in non relativistic Quantum Mechanics: rotations apart, the Galilei group acts by adding numerical multiples of the identity to the non commuting position and momentum operators .

In view of the Gel'fand--Naimark Theorem, the classical Minkowski Space \(M\) is described by the commutative C*-algebra of continuous functions vanishing at infinity on \(M\);  the classical coordinates can be viewed as commuting selfadjoint operators affiliated to that C*-algebras.

Similarly a {\itshape noncommutative} C*-algebra \(\mathscr E\) of Quantum Spacetime can be associated to the above relations. It was proposed in \cite{dopl1,dopl2}   by a procedure which applies to more general cases (see also Sections~\ref{subsec:basiccovariant} and~\ref{subsec:basic_model} below). 

Assuming that the \(q_\lambda,  Q_{\mu \nu}\) are selfadjoint operators and that the \(Q_{\mu \nu}\) 
commute {\itshape strongly} with one another and with the \(q_\lambda\), the relations above can be seen as a bundle of Lie algebra relations based on the joint spectrum of the \(Q_{\mu \nu}\).

We are interested only in representations which are regular in the sense that in their central decomposition only integrable representations of the corresponding Lie algebras appear. 

Such representations are described by representations of the group C*-algebra of the unique simply connected Lie group associated to the corresponding Lie algebra.

Hence the C*-algebra of Quantum Spacetime \(\mathscr E\) is the C*-algebra of a continuous field of group C*-algebras based on the spectrum of a commutative C*-algebra.

In our case,  that spectrum---the joint spectrum of the \(Q_{\mu \nu}\)---is the manifold \(\Sigma\) of the real valued antisymmetric 2-tensors fulfilling the same relations as the \(Q_{\mu \nu}\)  do: a homogeneous space of the proper orthochronous Lorentz group, identified with the coset space of \(SL(2,\mathbb{C})\) mod the subgroup of diagonal matrices.  Each of those tensors can be taken to its rest frame, where the electric and magnetic part are parallel unit vectors, by a boost specified by a third vector, orthogonal to those unit vectors; thus  \(\Sigma\) can be viewed as the tangent bundle to two copies of the unit sphere in 
3-space---its base  \(\Sigma_1\). 

The fibers, with the condition that \(I\) is not an independent generator but is represented by \(I\), are the C*-algebras of the Heisenberg relations in 2 degrees of freedom---the algebra of all compact operators on a fixed infinite dimensional separable Hilbert space.
 
The continuous field can be shown to be trivial, since it must contain a continuous field of one dimensional projectors---those corresponding to the orthogonal projection on the one dimensional subspace of multiples of the ground state vector for the harmonic oscillator (see \cite{dopl1}). 

The states whose central decomposition is supported by the base  \(\Sigma_1\), and for each point of the base correspond to the ground state for the harmonic oscillator, are precisely the states of {\itshape optimal localisation}, where the 
{\itshape sum} of the four squared uncertainties of the coordinates is minimal, and equal to \(2\) (see Section~\ref{subsec:ur_opt} below).

Thus the C*-algebra  of Quantum Spacetime \(\mathscr E\) is identified with the tensor product of
 the continuous functions vanishing at infinity on \(\Sigma\) and the algebra of compact operators.
 
 In the {\itshape classical limit} \(\lambda_P  \rightarrow 0\) the second factor deforms to the commutative C*-algebra of Minkowski space, but the first factor survives. When Quantum Spacetime is probed with optimally localised states its classical limit is \(M \times 
 \Sigma_1\), i.e. \(M\) acquires {\itshape compact extra dimensions}.
 
 Note that the mathematical generalisation of points are pure states, but only optimally localised pure states are physically appropriate.
 
 But to explore more thoroughly the Quantum Geometry of Quantum Spacetime we must consider {\itshape independent events}.
 
 Quantum mechanically \(n\) independent events ought to be described by the 
$n$-fold tensor product of \(\mathscr E\) with itself; considering arbitrary values on \(n\) we are led to use the direct sum over all \(n\).
 
 If \(A\) is the C*-algebra with unit over \(\mathbb{C}\), obtained 
adding the unit to \(\mathscr E\),  we 
will view the \((n+1)\) tensor power \(\Lambda_n (A)\) of \(A\) over \(\mathbb{C}\) 
as an \(A\)-bimodule with the product in \(A\), and the direct sum
\[ 
\Lambda (A)  = \bigoplus\limits_{n = 0}^\infty \Lambda _n (A) 
\] 
as the \(A\)-bimodule tensor algebra, where
\[
(a_1 \otimes a_2 \otimes . . . \otimes a_n) (b_1 \otimes b_2 \otimes . . . \otimes b_m) = a_1 \otimes a_2 \otimes . . . \otimes (a_n b_1) \otimes b_2 \otimes . . . \otimes b_m.
\] 
This is the natural ambient for the {\itshape universal differential calculus}, where the differential is given by
 \[ 
d (a_0 \otimes \dots \otimes a_n)  = \sum \limits_{k = 0}^n (-1)^k a_0 
\otimes \dots \otimes a_ {k-1} \otimes I \otimes a_k \otimes \dots \otimes a_n . 
\] 
As usual \(d\) is a graded differential, i.e., if 
\(\phi \in \Lambda (A), \psi \in \Lambda_n (A)\), we have 
\begin{align*} 
d^2 &= 0;\\
d (\phi \cdot \psi ) &= (d\phi ) 
\cdot \psi + (-1)^n \phi \cdot d \psi. 
\end{align*} Note that \(A = \Lambda_0 (A) 
\subset\Lambda (A)\), and the \(d\)-stable subalgebra \(\Omega (A)\) of \(\Lambda (A)\) generated by \(A\) is the { \itshape universal differential 
algebra}. In other words, it is the subalgebra generated by \(A\) and
\[
da  =  I \otimes a - a \otimes I
\]
as \(a\) varies in \(A\).

In the case of \(n\) independent events one is led to describe the spacetime coordinates of the \(j\)\nth event by \( q_j  =  I \otimes . . . I  \otimes \otimes q  \otimes I . . .  \otimes I\) (\(q\) in the \(j\)\nth place); in this way, the commutator between the different spacetime components of the \(q_j\) would depend on \(j\).

A better choice is to require that it does not; this is achieved as follows. The centre \(Z\) of the multiplier algebra of \(\mathscr E \) is the algebra of all bounded continuous functions on \(\Sigma\) with values in the complex numbers; so that \(\mathscr E\), and hence \(A\), is in an obvious way a  \(Z\)-bimodule. 

Therefore we can, and will, replace, in the definition of  \(\Lambda (A)\), the  \(\mathbb{C}\)-tensor product by the \(Z\)-bimodule-tensor product, so that
\[ 
dQ  =  0.
\]

As a consequence, the \(q_j\) and the \( 2^{-1/2}  (q_j  -  q_k)\), \(j\) different from \( k\), and \(2^{-1/2} dq\), obey the same spacetime commutation relations, as does the normalised barycenter coordinates,   \( n^{-1/2} (q_1 +  q_2 + . . . q_n)\); and the latter commutes with the difference coordinates.

These facts allow us to define a {\itshape quantum diagonal map} from \( \Lambda_n (A)\)  to \(A\), which leaves the functions of the barycenter coordinates alone, and evaluates on functions of the difference variables the {\itshape universal optimally localised map} which, when composed with a probability measure on \(\Sigma_1\), would give the generic optimally localised state (see Section~\ref{subsec:diagonal} below). 

Replacing the classical diagonal evaluation of a function of \(n\) arguments on Minkowski space by the quantum diagonal map allows us to define the {\itshape Quantum Wick Product} \cite{dopl5}. 

But working in  \(\Omega (A)\) as a subspace of \(\Lambda (A)\) allows us to use two structures \cite{dopl9}:
\begin{itemize}
\item the tensor algebra structure described above, where both the \(A\) bimodule and the \(Z\)   
  bimodule structures enter, essential for our reduced universal differential calculus;
\item the pre-C*-algebra structure of   \(\Lambda (A)\), which allows us to consider, for each 
  element \(a\) of  \(\Lambda_n (A)\), its modulus \((a^ {*}a)^ {1/2}\), its spectrum, and so on.\end{itemize}  
 In particular we can study the geometric operators: separation between two independent events,  area, 3-volume, 4-volume, given by
\begin{align*}
&dq,\\
&dq \wedge dq,\\
&dq \wedge dq \wedge dq,\\
&dq \wedge dq \wedge dq \wedge dq,
\end{align*} 
where, for instance, the latter is given by
\begin{align*}
V &=  dq \wedge dq \wedge dq \wedge dq =\\ 
&=\epsilon_{\mu\nu\rho\sigma}dq^{\mu}dq^{\nu}dq^{\rho}dq^{\sigma}.  
\end{align*}

Each of these forms has a number of spacetime components: e.g. \(4\) the first one (a vector), \(1\) the last one (a pseudoscalar).

It is found that, for each of those forms, each component is a normal operator, and that the sum of the square moduli of all spacetime components is bounded below by a multiple of the identity of unit order of magnitude. Although that sum is (except for the 4-volume!) not Lorentz invariant, the bound holds in any Lorentz frame (see Section~\ref{subsec:geometric} below).

In particular, the {\itshape Euclidean} distance between 
two independent events can be shown to have a lower bound of order one in 
Planck units. Two distinct points {\itshape can never merge to a point}.  
However, of 
course, the state where the minimum is achieved will depend upon the 
reference frame where the requirement is formulated. (The structure of 
length, area and volume operators on QST has been studied in full detail  
\cite{dopl9}.)

Thus the existence of a minimal length is not at all in contradiction 
with the Lorentz covariance of the model; note that models where the 
commutators of the coordinates are just numbers \(\theta\), which appear 
so often in the literature, arise as irreducible representations of our 
model; such models, taken for a fixed choice of \(\theta\) rather than for 
its full Lorentz orbit, necessarily break Lorentz covariance. To restore 
it as a twisted symmetry is essentially equivalent to going back to the 
model where the commutators are operators. This point has been recently 
clarified in great depth \cite{dopl6}.

On the other side, a theory with a fixed, numerical
commutator (a \(\theta\) in the sky; it could be hardly believed, but at least, in case, it ought to be thought in the CMB reference system, with respect to which we fly at a speed of \(600\) km per second!) can hardly be realistic.

The geometry of Quantum Spacetime and the free field theories on it are 
{\itshape fully Poincar\'e covariant}. The various formulation of interaction 
between 
fields, all equivalent on ordinary Minkowski space, provide inequivalent 
approaches on QST; but all of them, sooner or later, meet problems with 
{\itshape Lorentz covariance}, apparently due to the nontrivial action of 
the Lorentz group on the {\itshape centre} of the algebra of Quantum Spacetime. 
On this point in our opinion a deeper understanding is needed.

One can however introduce interactions in different ways, all preserving 
spacetime translation and space rotation covariance, that we discuss in Section 3; among these it is 
just worth mentioning here one of them, where one takes into account, in 
the very definition of Wick products, the fact that in our Quantum 
Spacetime \(n\) (larger or equal to two) distinct points can never merge to a point. But we can use the canonical quantum diagonal map mentioned above, which 
associates to 
functions of \(n\) independent points a function of a single point, evaluating a
conditional expectation which on functions of the differences takes a 
numerical value, associated with the minimum of the Euclidean distance (in 
a given Lorentz frame!).

The ``Quantum Wick Product''  obtained by this procedure leads to a 
perturbative Gell-Mann and Low formula free of ultraviolet divergences at each 
term of the perturbation expansion \cite{dopl5} . However, those terms have a 
meaning only after a sort of adiabatic cutoff: the coupling constant should be changed to a 
function of time, rapidly vanishing at infinity, say depending upon a cutoff time \(T\). But the limit \(T \rightarrow \infty\) is difficult problem, and there are indications it does not exist.

A major open problems is the following. Suppose we apply this construction to the normalised Lagrangean of a theory which is  renormalisable on the ordinary Minkowski space, with the counter terms defined by that ordinary theory, and with finite renormalization constants depending upon both the Planck length \(\lambda_P\) and the cutoff time \(T\). Can 
we find a natural dependence such that in the limit \(\lambda_P  \rightarrow 0\) and  \(T \rightarrow \infty\) we get back the ordinary renormalized Gell-Mann Low expansion on Minkowski space? This should depend upon a suitable way of performing a joint limit, which hopefully yields, for the physical value of  \(\lambda_P\), to a result  which is essentially independent of \(T\) within wide margins of variation, and can be taken as source of predictions to be compared with observations.

The common feature of all approaches is that, due to the quantum nature 
of spacetime at the Planck scale, {\itshape locality is broken} (even at the level 
of free fields, for explicit estimates see  \cite{dopl1}); 
in perturbation theory, its breakdown manifests itself in a non local kernel, which 
spreads the interaction 
vertices  \cite{dopl1,dopl4,dopl5} ; 
this forces on us the appropriate modifications of Feynman rules
\cite{dopl3}. 

However, it is worth noting that in Quantum Field Theory on the Minkowski space (and similarly on curved classical backgrounds) there are two aspects of locality.

First, the theory is defined by the {\itshape assignment} to bounded (nice) open regions in spacetime of algebras generated by the observables which can be measured within those regions.

Covariance is expressed by the fact that that assignment intertwines the actions of the spacetime symmetries on the regions and on the observables.

Second, that assignment should reflect Einstein causality: observables that are measured in regions between which no signal can be transmitted, ought to commute.

As we mentioned, the second assertion is bound to be lost if the gravitational forces \(between\) the elementary particles are taken into account.

But the first assertion, at least partially, can well be maintained.

Indeed, if we describe Minkowski space by the algebra of continuous functions vanishing at infinity, we can describe open sets through their characteristic functions, which are special selfadjoint idempotents in the Borel completion.

Similarly, a ``region'' in Quantum Spacetime can be described by a  selfadjoint idempotent \(E\) in the Borel completion of the C*-algebra of Quantum Spacetime.

To associate algebras of observables to such projections assume first that we wish to define on the basic model of Quantum Spacetime  the ordinary free field \(\phi\) over Minkowski space. 

The analogue of the von Neumann functional calculus on the $q_\mu$'s with functions whose Fourier transform is $L^1$ can be extended to operator valued distributions as Wightman fields (cf \cite{dopl1} and Section \ref{sec:basic_model} here below). This applies in particular to free fields.

The evaluation of \(\phi\) on the noncommuting operators \(q\) can be given by 
\begin{equation}\label{eq:phiq}
\phi(q)=\frac{1}{(2\pi)^{3/2}}\,\int(e^{iq_\mu k^\mu}\otimes a(\vec k)+e^{-iq_\mu k^\mu}
\otimes a(\vec k)^*)d\Omega^+_m(\vec k)
\end{equation}
where \(d\Omega^+_m(\vec k)=\frac{d^3\vec k}{2\sqrt{\vec k^2+m^2}}\) is the usual invariant measure over the  positive
energy hyperboloid of mass \(m\):
\[
\Omega^+_m=\{k\in\mathbb{R}^4\;/\;k_\mu k^\mu=m^2\ , k_0>0\}.
\]

This is an unbounded operator affiliated to the C*-tensor product \({\mathscr E}\otimes{\mathcal B}({\mathcal H})\), where \(\mathcal H\) is the Fock space.

Similarly, using the full Fourier transform of the field, any Wightman field on Minkowski space could be evaluated on \(\mathscr E\).

The free field defines a map from states \(\omega\in{\mathcal S}({\mathscr E})\) to operators on \({\mathcal H}\) by

\[\phi(\omega)\equiv\langle\omega\otimes id,\phi(q)\rangle\ ,\quad\omega\in{\mathcal S}({\mathscr E})\ .\]

The von Neumann algebra generated by bounded functions of these operators, as \(\omega\) varies in the set of states supported by \(E\), will be the {\itshape local algebra} \( \mathfrak{A} (E)\) associated to \(E\).

This map preserves inclusions and intertwines the actions of the Poincar\'e group, since the free field is covariant. The same would apply to any covariant field.

However, the local commutativity is lost, as well as the notion ``\(E\) is spacelike to \(F\)''.

The local algebras \(\mathfrak A(E)\) might show many unexpected behaviours. In the case of a free scalar neutral field, to a minimal \(E\) given by the product of the characteristic function of  a point in \(\Sigma_1\) with the spectral projection of the sum of squares of the coordinates associate to the interval \([0,2]\), we would get a commutative algebra; in the case of a free Dirac Field, a finite dimensional algebra. But spreading those algebras with spacetime translations in any tiny neighbourhood would lead to an irreducible algebra \cite{dopl15}. These results partly survive even for the scale invariant model of Quantum Spacetime with \(\lambda_P  =  0\) \cite{dopl16}.

Can we formulate an analogue of Locality as a sharp, physically compelling, principle, which reduces to ordinary locality at large scales?

The only way we can figure out to address this question relates to the Principle of local gauge invariance and of minimal form of the interactions.

In ordinary Field Theory these principles select {\itshape local point} interactions, and thus can be viewed as the root of locality.

We could speculate on the extension of those principles to Quantum Field Theory on Quantum Spacetime as the way to extend Locality.

But, unfortunately, already on Minkowski space those principle seem to have a crystal clear form only in classical field theory, and to be not amenable to any formulation in terms of local observables. And they seem to require anyway a formulation in terms of non observable quantities.

Hence at the moment we cannot say more than the fact that locality must break down on Quantum Spacetime.  

But  nonlocal effects 
should be visible only at Planck scales, and vanish fast for larger 
separations. If Lorentz invariance can be maintained by interactions, a 
point quite open at present, then we ought to expect that the analysis of the superselection structure, the notion of Statistics, conjugate sectors, the emergence of a compact group of global gauge symmetries, and even the Spin and 
Statistics Theorem, all deduced on the basis of the Principle of Locality, ought to remain true \cite{dopl17}.

That argument might, however, raise the objection that, in a theory which accounts 
for gravitational interactions as well, there might be no reasonable scattering 
theory at all, due to the well known paradox of loss of information, if black 
holes are created in a scattering process, destroying the unitarity of the S 
matrix.

Of course, this is an open problem; but one might well take the attitude that a 
final answer to it will come only from a complete theory, while at the moment we 
are rather relying on semiclassical arguments. Which might be quite a reasonable 
guide in order to 
get indications of local behaviours; but scattering theory involves the limit to 
infinite past/future times; and it might well be that interchanging these limits 
with those in which the semiclassical approximations are valid, or with the 
infinite volume limit in which the thermal behaviour of the vacuum for a 
uniformly accelerated observer becomes an exact mathematical statement, is 
dangerous, if not misleading. And whatever theory will account for Quantum 
Gravity, it should also describe the world of Local Quantum Field Theory as an 
appropriate approximation.
 
One might expect that a complete theory ought to be covariant under 
general coordinate transformations as well. This principle, however, is 
grounded on the conceptual experiment of the falling lift, which, in the 
classical theory, can be thought of as occupying an infinitesimal 
neighbourhood of a point.  In a quantum theory the size of a ``laboratory'' 
must be large compared with the Planck length, and this might pose 
limitations on general covariance.

On the other side elementary particle theory deals with collisions which 
take place in narrow space regions, studied irrespectively of the 
surrounding large scale mass distributions, which we might well think of 
as described by the vacuum, and worry only about the short scale effects 
of gravitational forces. 

We are thus lead to consider Quantum Minkowski Space as a more realistic 
geometric background for Elementary Particle Physics. But, as we briefly mentioned at the beginning,  the energy 
distribution in a generic quantum 
state will affect the Spacetime Uncertainty Relations, suggesting that the 
commutator between the coordinates ought to depend in turn on the metric 
field. 

Thus the spacetime commutation relations would become part of the {\itshape 
equations of motion}.

While in Classical General Relativity Geometry is part of the Dynamics, in this scenario also {\itshape Algebra} would be part of the Dynamics.

This might well be the clue to restore Lorentz covariance in the  theory of
interactions between fields on Quantum Spacetime.

On the other side, we mentioned how heuristic arguments suggest that the distance of acausal propagation of effects could increase near singularities.

This scenario could be related to the large scale thermal equilibrium 
of the cosmic microwave background (horizon problem). Actually, 
taking into account only of the Planck length as a universal lower bound for that distance of propagation, and assuming the simple model of a scalar massless field semiclassically interacting with the gravitational field (but treating EE exactly) shows that the \emph{effect} of the divergence of the minimal distance of acausal propagation shows up, solving the horizon problem without any inflationary hypothesis.

Similarly one could wonder whether the non vanishing of the 
Cosmological Constant  is related to the dependence of the commutators of the coordinates upon the metric \cite{dopl7}. And to the fact that noncommutativity at the Planck scale might manifest itself as an effective repulsion; in which case it might well be an explanation of an inflationary potential.

\section{The basic model: an example of Quantum Geometry} 
\label{sec:basic_model}
\subsection{The basic model and its covariant representations}
\label{subsec:basiccovariant}
The basic model arises from the simplifying {\itshape ansatz} 
that the commutators
\(Q^{\mu\nu}=-i[q^\mu,q^\nu]\) are central, namely they strongly
commute with the coordinates \(q^\mu\). To fix domain ambiguities and select
reasonably regular representations, we understand the formal 
definition of the antisymmetric 2-tensor \(Q^{\mu\nu}\) as a reminder
of the Weyl relations
\begin{equation}\label{eq:weyl_rels} 
e^{ih_\mu q^\mu}e^{ik_\nu q^\nu}=e^{-\frac i2h^\mu Q^{\mu\nu} h\nu}e^{i(h+k)_\mu q^\mu},\quad h,k\in\mathbb R^3,
\end{equation}
where we took care of using the Lorentz metric to parametrise the 4-parameters 
group \(k\mapsto e^{ikq}=e^{ik_\mu q^\mu}\). In what follows, formal commutation rules
will always be understood  as shorthands for the regular Weyl form.

As described in Section~\ref{sec:why}, covariant quantitative conditions on the 
commutators amount to make a choice of the quantities \(a,b\) of the
two independent ``scalars'' which can be formed out of an antisymmetric tensor: 
\[
Q^{\mu\nu}Q_{\mu\nu}=aI,\quad  \left(\frac{1}{4}Q^{\mu\nu}(\star Q)_{\mu\nu}\right)^2=bI.
\]
The choice \(a=0,b=1\)---which in a sense is the most symmetric, 
see \cite{dopl1}---results in Heisenberg-like uncertainty relations which
have the same form as the desired heuristically motivated relations \eqref{eq:stur}.

A first, {\itshape a priori} only partial classification of the irreducible 
representations is provided by the remark that, by the Schur lemma, the \(2\) tensor of the commutators must be of the form \(i\sigma^{\mu\nu}\) for some
constant real 
antisymmetric \(2\)-tensor \(\sigma=(\sigma^{\mu\nu})\).
It follows from the quantisation conditions that such a 
\(\sigma\) should fulfil
\[
\sigma^{\mu\nu}\sigma_{\mu\nu}=0,\quad  \left(\frac{1}{4}\sigma^{\mu\nu}(\star \sigma)_{\mu\nu}\right)^2=1.
\]
Let \(\Sigma\) be the manifold of all antisymmetric \(2\)-tensors 
fulfilling the above conditions; it is by construction a homogeneous space
under the natural action \(\sigma\mapsto{}_\varLambda\sigma=\varLambda\sigma\varLambda^t=( {\varLambda^{\mu}}_{\mu'}\sigma^{\mu'\nu'}{\varLambda^{\nu}}_{\nu'})\)
of the full Lorentz group.

Therefore, in order to classify all irreducible representations, it  is sufficient to classify all the equivalence classes of irreducible regular representations with commutators which are multiples of the identity.

We next observe that there is a natural choice for \(\sigma_0\):
the standard symplectic matrix 
\[
\sigma_0=\begin{pmatrix}0&0&1&0\\0&0&0&1\\-1&0&0&0\\0&-1&0&0\end{pmatrix}\in\Sigma.
\]
Upon renaming 
\[
(P_1,P_2,Q_1,Q_2):=(q_{\sigma_0}^0,q_{\sigma_0}^1,q_{\sigma_0}^2,q_{\sigma_0}^3),
\]
the relations
\[
[q^\mu_0,q^\nu_0]=i\sigma_0^{\mu\nu}
\]
take the form of
the Canonical Commutation Relations
\[
[P_j,Q_k]=-i\delta_{jk},\quad [P_j,P_k]=[Q_j,Q_k]=0,
\]
for two canonical pairs \((P_1,Q_1)\) and \((P_2,Q_2)\).

This fact---which of course must be regarded solely as a mathematical identification without any {\itshape direct} physical interpretation---is very lucky, as it completely
solves the classification problem for irreducible representations of our
spacetime relations, by reducing it to von Neumann uniqueness: there is
only one irreducible representation 
\begin{equation}\label{eq:basic_rep}
q_0:=q_{\sigma_0}=(P_1,P_2,Q_1,Q_2)
\end{equation} 
with commutators \(i\sigma_0I\), up to equivalence; where \(P_j,Q_j\) 
are canonical Schr\"odinger operators. 

According to the previous remark, it follows that for every \(\sigma\in\Sigma\) there is one and one only regular irreducible representation \(q_\sigma=\varLambda_\sigma q_0\)
with commutators \(i\sigma = i\varLambda_\sigma \sigma_0 \varLambda_\sigma^t\), up to equivalence. 

The manifold \(\Sigma\) may be identified with the quotient \(\mathscr L/G_0\)
of the full Lorentz group by the stabiliser of \(\sigma_0\), which provides
the possibility of building \(\sigma\mapsto\varLambda_\sigma\) as a Borel 
section. Hence we have a complete classification of the representation theory
of the spacetime commutation relations.

Not only the occurrence of the standard symplectic matrix \(\sigma_0\) 
in \(\Sigma\) is lucky; it also is fascinating, for two quantum models
with quite distant underlying physical motivations and interpretation---the non relativistic 
quantum mechanics of a material point on the plane
and the gravity-induced (semiclassical) quantisation of the Minkowski 
spacetime---both 
rely on the very same basic building blocks: canonical pairs of Schr\"odinger 
operators. We also observe that 
the whole argument would have failed if the dimension of spacetime were odd,
precisely because the canonical operators come in pairs.

Next we address the question whether there is a representation \(q^\mu\) 
which is  Lorentz covariant, in the precise sense that there is 
a strongly continuous unitary representation \(U\) of the Lorentz group 
on the representation Hilbert space \(\mathscr H_q\) such that
\[
U(\varLambda)^*q^\mu U(\varLambda)={\varLambda^\mu}_{\mu'} q^{\mu'},
\]
where the closure of the operator on the right is implicitly understood,
or equivalently we regard the above as a shorthand of the corresponding
transformation of the Weyl operators.

Correspondingly,
\[
U(\varLambda)^*Q^{\mu\nu} U(\varLambda)={\varLambda^\mu}_{\mu'}
{\varLambda^\nu}_{\nu'} Q^{\mu'\nu'},
\]
which prevents the possibility for a covariant representation 
to be irreducible; on the contrary it will have to be highly reducible.

For every representation \(q\) of the relations \eqref{eq:weyl_rels}, 
the joint spectrum \(\jSp(Q)\) of the 16 operators \(Q^{\mu\nu}\) may be 
regarded as a manifold of antisymmetric real tensors 
\(\sigma=(\sigma^{\mu\nu})\), namely a submanifold of \(\Sigma\). 
If \(q^\mu\) is a covariant representation in the sense of above, 
necessarily \(\jSp(Q)\) is a homogeneous space under the Lorentz action;
hence it must coincide with the whole \(\Sigma\):
\[
q^\mu\text{\ covariant} \quad\Rightarrow \jSp(Q)=\Sigma.
\]

As a consequence, a covariant representation must weakly contain at least one 
representative \(q_\sigma\) for every \(\sigma\in\Sigma\). 

To construct a covariant representation, it would be sufficient to use
a quasi-invariant regular positive measure. However,
such a measure can be chosen to be even invariant:
we may use the  
projection map \(\mathscr L\mapsto\mathscr L/G_0=\Sigma\) 
and the Haar measure on \(\mathscr L\). Hence we take
the Hilbert space 
\[
\mathscr H_q=L^2(\mathscr L,\mathfrak H)
\]
of square summable, \(\mathfrak H\)-valued functions of \(\mathscr L\),
where \(\mathfrak H\) is the Hilbert space on which the Schr\"odinger operators
\(P_1,P_2,Q_1,Q_2\) act. Using the basic representation \eqref{eq:basic_rep},
we may set
\begin{align}
(q^\mu\Psi)(M)={M^\mu}_\nu q^\nu_0\Psi(M),\quad \Psi\in \mathscr D(q^\mu),\\
(U(\varLambda)\Psi)(M)=\Psi(\varLambda^{-1}M),\quad \Psi\in \mathscr H_q.
\end{align}
If we choose the Schr\"odinger representation 
\(P_j=-i\partial/\partial s_j,Q_j=s_j\cdot\) on \(\mathfrak H=L^2(\mathbb R^2,d^2s)\),
then \(\mathscr H_q\simeq L^2(\mathscr L\times\mathbb R^2,d\varLambda\,d^2s)\), and the operators
\(q^\mu\) are essentially selfadjoint e.g.~on the smooth, compactly supported
functions of \(\mathscr L\times\mathbb R^2\). 
Every other covariant representation is quasi-equivalent to the above.

The problem of obtaining a Poincar\'e covariant representation is easily
solved by doubling the underlying Schr\"odinger pairs, see \cite{dopl1}

\subsection{Uncertainty relations and optimal localisation}
\label{subsec:ur_opt}
It is convenient to identify the antisymmetric \(4\times 4\) 
matrices 
with the pairs \((\vec e,\vec m)\) of  their ``electric''
and ``magnetic'' parts of
\[
\begin{pmatrix}0&e_1&e_2&e_3\\-e_1&0&m_3&-m_2\\-e_2&-m_3&0&m_1\\-e_3&m_2&-m_1&0\end{pmatrix}
\]
One easily checks that, if \(\sigma=(\sigma^{\mu\nu})=(\vec e,\vec m)\),
then \((\sigma_{\mu\nu})=(-\vec e,\vec m)\) and \((\star\sigma_{\mu\nu})=
(-\vec m,\vec e)\). Moreover,
\[
\sigma=(\vec e,\vec m),\tau=(\vec f,\vec n)\quad\Longrightarrow\quad
\sigma^{\mu\nu}\tau_{\mu\nu}=2(\vec m\cdot\vec n-\vec e\cdot \vec f).
\]
With these notations,
\[
\Sigma=\{(\vec e,\vec m):\vec e\cdot\vec m=\pm 1,\vnorm{\vec e}=\vnorm{\vec m}\},
\]
where \(\vnorm{\cdot}\) is the Euclidean length; moreover,
\[
(\vec e,\vec m)\in\Sigma\quad\Rightarrow\quad \vnorm{\vec e}\geqslant 1,\vnorm{\vec m}\geqslant 1;
\]
this fact will be important in the derivation of the uncertainty relations.
Note also that the standard symplectic matrix corresponds to the second
vector of the canonical bases \(\{\vec n_1,\vec n_2,\vec n_3\}\) of
\(\mathbb R^3\): \(\sigma_0=(\vec n_2,-\vec n_2)\). 

If \(\varLambda=(\begin{smallmatrix}1&0\\0&R\end{smallmatrix})\) for \(R\in O(3,\mathbb R)\), then \(\varLambda (\vec e,\vec m)\varLambda^t=(R\vec e,\pm R\vec m)\) where \(\pm\det R=1\). The only subset of \(\Sigma\) which is
invariant under orthogonal transformations is
\[
\Sigma^{(1)}=\{(\vec e,\pm \vec e):\vnorm{\vec e}=1\}
\]
which has two connected components \(\Sigma^{(1)}_\pm\), both evidently
isomorphic to the \(2\)-sphere \(S^2\). It follows (cf introduction) 
that \(\Sigma\) itself
has two connected components \(\Sigma_\pm\), each of which is isomorphic with the tangent space of \(S^2\).

We may now sketch the argument by which the uncertainty relations 
\eqref{eq:stur} follow 
from the quantisation conditions; it is sufficient to prove 
\eqref{eq:stur}  
for every irreducible \(q_\sigma\), and for every vector state 
\(\omega(\cdot)=(\psi,\cdot\psi)\), \(\psi\in\mathscr D(q_\sigma)\)). 
With \(\sigma=(\vec e,\vec m)\in\Sigma\), 
the (generalised) Heisenberg uncertainty theorem gives
\[
\Delta_\omega(q_\sigma^0)\Delta_\omega(q_\sigma^j)\geqslant
\frac12\omega(|[q_\sigma^0,q_\sigma^j]|)=
\frac12|e_j|;
\]
\eqref{eq:stur_electric} then follows from
\(1\leqslant \vnorm{\vec e}=\left(\sum_j|e_j|^2\right)^{1/2}\leqslant\sum_j|e_j|\).
A similar argument (using \(|\vec m|\geqslant 1\)) gives \eqref{eq:stur_magnetic}.

The non-invariant 
quantity \(\sum_\mu\Delta(q_\mu)^2\)
provides information about the localisation properties of a state
according to a given observer. Given a state \(\omega\) on an
irreducible representation \(q_\sigma\), we have
\[
\sum_\mu\Delta_\omega(q_\sigma^\mu)^2\geqslant \sqrt{2+|\vec e|^2+|\vec m|^2},
\]
where \(\sigma=(\vec e,\vec m)\in\Sigma\), and provided \(\omega\) is in the domain of the involved operators (see \cite[Prop. 3.4]{dopl1} for more details).

Two questions arise:
\begin{enumerate}
\item given any \(\sigma\in\Sigma\), do states \(\omega\) on \(q_\sigma\) 
exist, such that the above bound is attained?
\item the bound itself is minimal when \(\sigma\in\Sigma^{(1)}\), in which case it becomes
\[
\sum_\mu\Delta_\omega(q_\sigma^\mu)^2\geqslant 2,\quad\sigma\in\Sigma^{(1)};
\]
do states \(\omega\) on \(q_\sigma\) for \(\sigma\in\Sigma^{(1)}\)
exist, such that the above bound is attained?
\end{enumerate}
While the answer to the general question 1) is unknown, question 2) is easy
to deal with. If \(\sigma\in\Sigma^{(1)}\), then \(\sigma=\varLambda\sigma_0\varLambda^t\) for some 
\(\varLambda=(\begin{smallmatrix}1&0\\0&R\end{smallmatrix})\), where
\(R=(R_{jk})\in O(3,\mathbb R)\). Then \(q_\sigma^j=\sum_{k=1}^3R_{jk}q_0^k\)
and
\begin{align*}
\sum_\mu{q_\sigma^\mu}^2&={q_\sigma^0}^2+\sum_{j=1}^3{q_\sigma^j}^2=
\sum_\mu{q_0^\mu}^2=\\
&=P_1^2+P_2^2+Q_1^2+Q_2^2,
\end{align*}
namely twice the Hamiltonian of the harmonic oscillator on the plane; 
the optimal localisation states are precisely the translates of the 
ground state of the harmonic oscillator (the canonical coherent states) and
\(\sum_\mu\Delta(q_\sigma^\mu)^2\geqslant 2\).

If instead we work with a state \(\omega\) on the fully covariant 
representation \(q\), define the probability measure \(\mu_\omega\)
on \(\Sigma\) by
\(f\mapsto \omega(f(Q))=\int_\Sigma f(\sigma)d\mu_\omega(\sigma)\), where
\(f(Q)\) is the joint bounded continuous functional calculus of the \(Q^{\mu\nu}\)'s. If \(\omega\) is in the domain of all \(q^\mu,{q^\mu}^2\), 
then 
\[
\sum_\mu\Delta_\omega(q^\mu)^2\geqslant \int_\Sigma d\mu_\omega(\sigma)\sqrt{2+|\vec e_\sigma|^2+|\vec m_\sigma|^2},
\]
where \((\vec e_\sigma,\vec m_\sigma)=\sigma\). 
Hence the lower bound becomes 
\[
\sum_\mu\Delta_\omega(q^\mu)^2\geqslant 2
\]
which is attained if \(\mu_\omega\) has support in \(\Sigma_1\) and
\(\omega\) acts as a superposition of canonical coherent states on each \(q_\sigma\) contained in \(q\), with \(\sigma\in\Sigma_1\); we shall make this
more transparent in the next section.

\subsection{The C*-algebra of the basic model}
\label{subsec:basic_model}
It is intuitively clear that we face a trivial bundle structure over 
\(\Sigma\): over each \(\sigma\in\Sigma\) there is a CCR-Weyl algebra, so that
the universal C*-algebra to which every regular representation of the Weyl relations is affiliated is
\[
\mathscr E=\mathbb C_0(\Sigma,\mathcal K)\simeq \mathcal C_0(\Sigma)\otimes\mathcal K),
\]  
namely the trivial continuous field of C*-algebras over \(\Sigma\) with standard fibre \(\mathcal K\), the compact operators over the separable, infinite dimensional Hilbert space \(\mathfrak H\). The multipliers C*-algebra \(M(\mathscr E)\) is easily identified with \(\mathcal C_b(\Sigma,B(\mathfrak H)\). 

While we
refer to \cite{dopl1} for the details of the proof why that bundle is trivial, 
we shall describe here how to work with
this algebra. 

We follow Weyl's prescription for quantisation:
\[
f(q)=\int dk \hat f(k)e^{ik_\mu q^\mu},\quad f\in L^1(\mathbb R^4)\cap\widehat{L^1(\mathbb R^4)},
\]
where \(\hat f(k)=(4\pi)^{-4}\int dx\,f(x)e^{ik_\mu x^\mu}\)  is the usual Fourier 
transform; in practice, the idea is to replace the usual
plane waves which build \(f\) up with their quantised counterpart, the Weyl 
operators. 

Since the commutators are not multiples of the identity, a product
\(f(q)g(q)\) is not of the form \(h(q)\); the Weyl-quantised functions do not close to an algebra of operators.

To circumvent this, we enlarge the class of functions to be quantised. 
We consider 
functions \(f(\sigma;x)\) of both \(\sigma,x\)
as elements 
of \(\mathcal C_0(\Sigma,L^1(\mathbb R^4))\), the space of continuous 
\(L^1(\mathbb R^4)\) valued functions
of \(\Sigma\), vanishing at infinity. For each \(\sigma\) define
\(\hat f(\sigma;k)=(4\pi)^{-4}\int dx\,f(\sigma;x)e^{ik_\mu x^\mu}\).

Whenever both \(f,\hat f\) are in \(\mathcal C_0(\Sigma,L^1(\mathbb R^4))\)---in which case we call \(f\) a {\itshape symbol}---we may construct the operator
\(f(Q;q)\), where the \(Q\) dependence is understood in the sense of joint
functional calculus, and the \(q\) dependence in the sense of Weyl 
quantisation. In more detail, 
if \(Q^{\mu\nu}=\int_\Sigma\sigma^{\mu\nu} dE(\sigma)\) is the joint
spectral resolution of the \(Q^{\mu\nu}\)'s,
\[
f(Q,q)=\int_\Sigma dE(\sigma)\int_{\mathbb R^4} dk \hat f(\sigma,k)e^{ikq},
\]
which is unambiguous, since the Weyl operators \(e^{ikq}\) and the joint 
spectral projections of the \(Q^{\mu\nu}\)'s commute.

A short computation with the Weyl relations gives the generalised symbolic
calculus, defined as the pull-back of the operator product to symbols:
\[
f(Q;q)g(Q;q)=(f\star g)(Q;q),
\]
where the \(\star\)-product 
\begin{equation}
(f\star g)(\sigma;x)=\frac1{(2\pi)^{4}}
\int da\int db\, f(\sigma;a)g(\sigma;b)e^{2i(a-x)_\mu\sigma^{\mu\nu}(b-x)_\nu}.
\end{equation}
may be regarded as a field of
\(\star_\sigma\)-products over \(\Sigma\):
\[
(f\star g)(\sigma;\cdot)=f(\sigma,\cdot)\star_\sigma g(\sigma;\cdot)
\]
Moreover, \(f(Q,q)^*=\bar f(Q;q)\). We thus equipped the space 
\(\mathscr S(\Sigma)\) of symbols with a product and an involution which make 
it a *-algebra, since they inherit all the relevant properties (associativity, involutivity,\ldots) from being 
the pull-back of the operator product and involution; it may be turned into
a Banach *-algebra taking its completion under the norm \(\|f\|=\sup_\sigma\|\hat f(\sigma,\cdot)\|_{L^1}\), with universal enveloping C*-algebra \(\mathscr E\). 
The algebra \(\mathscr S(\Sigma)\) of symbols may 
be regarded as an algebra of continuous sections for \(\mathscr E\).
Note that, if \(q\) is the fully covariant representation, 
\(f\mapsto f(Q;q)\) defines a faithful, covariant 
representation of \(\mathscr S(\Sigma)\):
\[
U(a,\varLambda)f(Q;q)U(a,\varLambda)=f(\varLambda^{-1}Q{\varLambda^{-1}}^t,\varLambda^{-1}(q-aI)),\quad (a,\varLambda)\in\mathscr P,
\] 
which extends to a faithful covariant
representation of \((\mathscr E,\alpha)\), where the action \(\alpha\) of the Poincar\'e group is the normal extension of the natural action on symbols. 
Hence we have an essentially unique
covariant representation of the C*-dynamical system \((\mathscr E,\alpha)\). We
thus feel free to understand \(f(Q;q)\) as indicating  
equivalently an operator 
(a represented element of the algebra), a symbol, or an abstract element of 
the algebra.

Let \(\omega\) is a state on \(q\) with optimal localisation and expectations
\(\omega(q^\mu)=a^\mu\). If \(\mu_\omega\) is the
associated measure on \(\Sigma\)---supported by \(\Sigma_1\)---as described at the end of Section \ref{subsec:ur_opt}, we have
\[
\omega(f(Q;q))=\int_{\Sigma_1}d\mu_\omega(\sigma)(\eta_a f)(\sigma)
\]
where for each localisation centre \(a\) the localisation map 
\(\eta_a :M(\mathscr E)\rightarrow\mathcal C(\Sigma_1)\)
is defined by
\[
(\eta_a f)(\sigma)=\int\limits_{{\mathbb R}^4}dk\hat f(\sigma;k)
e^{ika-{|k^2|/2}}
\]
and normal extension. It may be convenient to define \(\mathscr E_1=\mathcal C(\Sigma_1,\mathcal K)\), \(Z_1=\mathcal C(\Sigma_1)\) the centre of \(M(\mathscr E_1)\) and \(\eta_{a,1}=\eta_a\restriction_{\Sigma_1}\) 
as the restriction of \(\eta_a\) to \(\mathscr E_1\). 
Hence \(\eta_{a,1}\) is a morphism of \(Z_1\)-modules, and \(\eta_a\) is a conditional expectation, in  a natural way.

\subsection{Many events and the diagonal map}
\label{subsec:diagonal}
In order to develop a Quantum Geometry, we must identify the 
coordinates of multi-events. Since we want them to be independent, the 
usual prescription is to take tensor products: we regard each set
\begin{equation}
q_j^\mu=I^{\otimes(j-1)}
\otimes q^\mu\otimes I^{\otimes(n-j-1)}, \qquad \mu=0,\dots,3,
\end{equation}
as the coordinates of the \(j\)\nth event.

Then a segment may be identified by its two independent endpoints \(q_j,q_k\),
or even better with the separation operator \(q_j-q_k\). 

Since the theory is covariant under translations, we should expect
the separations \(q_j-q_k\) of two events
to be statistically independent from the average position 
\[
\bar q=\frac1n\sum_{j=1}^nq_j
\]
of all the \(n\) events.
We immediately check that
\[
[\bar q^\mu,(q_j-q_k)^\nu]=\frac1n(Q_j^{\mu\nu}-Q_k^{\mu\nu}),
\]
where \(Q_j^{\mu\nu}=-i[q_j^\mu,q_j^\nu]\); which does not vanish if
\(\otimes\) is understood as the tensor product of complex spaces. 
This forces us to understand \(\otimes\) as the tensor product
\(\otimes_Z\) of \(Z\)-modules, where \(Z\simeq\mathcal C_b(\Sigma)\)
is the centre of the multiplier algebra \(M(\mathscr E)\); intuitively,
this amounts to take the usual tensor product fibrewise on \(\Sigma\) . 
Since \(Q\) is affiliated to \(Z\), with this
position 
\[
Q_j=Q, \quad j=1,\dotsc,n,
\]
and
\begin{align}\label{eq:comm_many_events}
[q_j^\mu,q_k^\nu]&=iQ^{\mu\nu},\\
[\bar q^\mu,(q_j-q_k)^\nu]&=0, \quad j,k=1,\dotsc,n.
\end{align}
The coordinates \(q_j^\mu\) are then 
affiliated with the C*-algebra\footnote{Of course
\(\mathcal K^{\otimes n}\simeq \mathcal K\), so that \(\mathscr E^{(n)}\simeq
\mathscr E\).} 
\[
\mathscr E^{(n)}=\underbrace{\mathscr E\otimes_Z\dotsm\otimes_Z\mathscr E}_{\text{\(n\) factors}} \simeq
\mathcal C_0(\Sigma, \underbrace{\mathcal K\otimes\dotsm\otimes\mathcal K}_{\text{\footnotesize \(n\) factors}}).
\]

We now are in condition to construct a natural (non surjective) 
*-monomorphism from
\(M(\mathscr E^{(n)})\) to \(M(\mathscr E^{(n+1)})\).

By construction
\[
\bar Q^{\mu\nu}:=-i[\bar q^\mu,\bar q^\nu]=-\frac inQ^{\mu\nu};
\]
namely the same commutation relations of the basic model, with the Planck 
length~1 replaced by \(\sqrt{1/n}\)  
(in natural units where \(\lambda_P=1\)). It follows that \(\bar q\) is an amplification of \(q/\sqrt n\). Moreover we have the identity
\[
q^j=\bar q+\frac1n\sum_k(q_k-q_j).
\] 

The commutation relations \eqref{eq:comm_many_events} may 
be equivalently realised by taking
\[
\bq_j^\mu=\frac1{\sqrt n}q^\mu\otimes_Z I^{\otimes n}+\frac1n I\otimes_Z\sum_k(q_k-q_j),
\]
so that
\[
[\bq_j^\mu,\bq_k^\nu]=i\delta_{jk}Q^{\mu\nu}
\]
and we recognise that \(\bar \bq=\frac1{\sqrt n}q\otimes_Z I^{\otimes n}\) 
and \(\bq_j-\bq_k=I\otimes_Z(q_j-q_k)\) live in different tensor factors.

It follows that, with
\[
f(Q;q_1,\dotsc,q_n)=\int dk_1\dotsm dk_n\hat f(Q;k_1,\dotsc,k_n)e^{i(k_1q_1+\dots+k_nq_n)},
\]
the map
\begin{align*}
\beta:f(Q;&q_1,\dotsc,q_n)\mapsto f(Q;\bq_1,\dotsc,\bq_n)=\\
&=\int dk_1\dotsm dk_n\hat f(Q;k_1,\dotsc,k_n)e^{\frac{i}{\sqrt n}\sum_jk_jq}\otimes_Z
e^{\frac in\sum_{jh}k_j(q_h-q_j)}
\end{align*}
extends to the announced *-monomorphism. This is interesting because it provides
a tensor separation between the average position of a family of \(n\) independent events, and the algebra of the relative positions. This suggests to set
the relative positions as close to zero as possible, compatibly with positivity
in the algebra, leaving a function of the average position  
(and the centre) alone, to be understood as a noncommutative analogue of 
the classical evaluation of a function \(f(x_1,\dotsc,x_n)\) 
at \(x_1=x_2=\dotsb=x_2\).

Now, let \(\omega_a\) be an optimally localised state with localisation centre \(a\) and associated measure \(\mu_\omega\) on \(\Sigma_1\); the idea we have in mind is to compose the above *-monomorphism with 
``\(\text{id}\otimes\omega_a\otimes\dotsm\otimes\omega_a\)'', so to set the separations 
\(q_j-q_k\) to their minima, while leaving a function of \(\bar q\) alone. However \(\omega_a\)
is not a \(Z\)-module map, hence such a tensor product is not well defined.  
Taking seriously that the centre should be regarded as a point independent 
background, and recalling from the end of subsection \ref{subsec:basic_model} 
that \(\omega_a=\int d\mu_\omega(\sigma)\circ\eta_a\)
and \(\eta_{a,1}=\eta_a\circ\restriction_{\Sigma_1}\),
we may define the desired quantum diagonal map  \(E^{(n)}\) as 
\[
\mathcal C_b(\Sigma,\mathcal K^{\otimes n})\overset{\beta}{\rightarrow}
\mathcal C_b(\Sigma,\mathcal K^{\otimes(n+1)})\overset{\restriction_{\Sigma_1}}{\rightarrow}
\mathcal C(\Sigma_1,\mathcal K^{\otimes(n+1)})\overset{\Phi}{\rightarrow}\mathcal C(\Sigma_1,\mathcal K)
\]
where 
\[\Phi=\text{id}\otimes_{Z_1}\underbrace{\eta_{a,1}\otimes_{Z_1}\dotsm\otimes_{Z_1}\eta_{a,1}}_{\text{\footnotesize{\(n\) factors}}}.\] 
It is an obvious consequence of translation covariance that the resulting map
does not depend on the choice of \(a\). We find
\begin{align}\label{eq:diagonal}\nonumber
E^{(n)}&f(Q;q_1,\dotsc,q_n)=\\&=\int da_1\dotsm da_n 
e^{-\frac12\sum_j|a_j|^2}\delta^{(4)}\Big(\frac1n\sum_ja_j\Big)
f(Q;\bar q+a_1,\dotsc,\bar q+a_n),
\end{align}
where \(\bar q\) now are the coordinates with characteristic length \(\sqrt{1/n}\) and affiliated to \(\mathcal C(\Sigma_1,\mathcal K)\), while
\(|a|^2=\sum_\mu (a_\mu)^2\).

The map so constructed is naturally covariant under orthogonal transformations,
but not under Lorentz boosts.

\subsection{Planckian bounds on geometric operators}
\label{subsec:geometric}
The choice of the \(Z\)-module tensor product to form coordinates
of many events, discussed in the preceding section, was motivated by
the necessity that \([q^\mu,q^\nu]\otimes_Z I-I\otimes_Z[q^{\mu'},q^{\nu'}]=0\) which, in the universal differential calculus, reads 
\[dQ=0.\] 

With \(q_j=I^{\otimes_Z(j-1)}\otimes_Z q\otimes I^{\otimes_Z(n-j-1)}\) the coordinates of the \(j\)\nth of \(n\) events, 
\[
dq_j=q_{j+1}-q_j,\quad j=1,\dotsc,n-1
\] is the separation between
2 of \(n\) events.

The operator \(\sum_\mu {q^\mu}^2\) may be regarded as the square 
Euclidean distance between the event and the (classical) origin, 
and thus has no direct physical interpretation; we already observed that it 
is bounded below by 2. 
More interesting is the Euclidean distance \(\sum_\mu {dq^\mu}^2\)
between two events. We easily compute
\[
[dq^\mu,dq^\nu]=2iQ^{\mu\nu},
\]
namely the same commutation relations as the basic coordinates, with characteristic length \(\sqrt 2\) (or \(\sqrt 2\lambda_P\), in generic units). 
It follows that the same bound on the square Euclidean length of \(q\)---appropriately scaled---holds true for the square Euclidean length of \(dq\):
\[
\sum_\mu {dq^\mu}^2\geqslant 4.
\]
While observers connected by a Lorentz boost will disagree in general
about the localisation 
states where this bound
can be attained, they agree on the bound itself, which thus is a quantity with an 
invariant meaning and a physical interpretation, and may be experimentally 
tested (at least in principle). This shows that a fully covariant theory may 
well
be characterised by two distinct physically meaningful invariant quantities---the light speed and the Planck length---{\itshape without
any contradiction with the Lorentz-Fitzgerald contraction}. In a sense, Special 
Relativity already is ``Doubly-Special'' in the sense of \cite{AC}, {\itshape without any modification 
(deformation) of the Lorentz action.}

This is already an interesting geometric bound, though very elementary; 
by the way, it provides a clear example why a minimal length needs not being 
realised  
as a limitation on the precision which can be attained when 
measuring a 
single coordinate\footnote{Such a limitation could not be obtained in any case
if coordinates have to be represented by selfadjoint operators, unless 
the availability of (generalised) eigenstates is restricted.}, nor by requiring a discrete spectrum (in this model---as well as in any translation-invariant model---the spectrum of the coordinates is continuous).

In \cite{dopl9}, the spectra of 2, 3 and 4-volume operators mentioned in Section~\ref{sec:why} are discussed in some detail, for the case of the coordinates of the basic model. Note that, in the 
definition of such ``quantum form-operators'' operators, the order of products 
does matter, so that they are not, and cannot be, (essentially) selfadjoint. 
The presence of a non trivial polar decomposition may be regarded as a quantum 
generalisation of the classical notion of orientation.
However---quite surprisingly---it is possible to show that they all are normal, so that they have a well defined spectral theory.

The findings of \cite{dopl9} are
\begin{enumerate}
\item the square Euclidean length of the separation \(dq\) between two independent events is bounded below by 4; its square Lorentzian length has continuous spectrum, pure Lebesgue, including the whole real line;
\item the sum of the squares of the components of both the space-time and space-space area operators \(dq^0\wedge dq^j\) and
\(dq^j\wedge dq^k\) have spectral values with absolute value bounded below by 1;
\item the 4-vector \(V^\mu=\bigwedge_{\nu\neq\mu}dq^\nu\) whose components are the 3-volume operators has Euclidean length bounded below by 8;
its time component alone has spectrum \(\mathbb C\);
\item the 4-volume operator \(V\) has spectrum
\[
\sigma(V)=\pm 2+\sqrt{5}\mathbb Z+i\mathbb R,
\]  
whose distance from \(0\) is \(\sqrt 5-2\).
\end{enumerate}
Apart the numeric factors, all bounds on \(n\)-volume operators 
(above expressed in natural units where
\(\lambda_P=1\)) are of order \(\lambda_P^n\), 
consistently with their physical 
dimensions.

\section{Quantum Field Theory on Quantum Spacetime: the various approaches and their problems} 
\label{sec:QFT}
The problem of a Quantum Field Theory of Gravitation, eighty years after the pioneering paper by M.~P.~Bronstein on the quantum {\it {linearized}} Einstein theory~\cite{dopl14}, is still open. 

It is therefore not entirely surprising if, twenty years after the publication of~\cite{dopl1}, the study of the interactions between quantum fields on Quantum Spacetime remains somewhat unsatisfactory.  For, even if very simple forms of interactions are studied, the underlying geometry keeps into account some quantum aspects of gravitation near singular regimes. 

While a large number of calculations have been performed and some conceptual issues have been raised, leading to a better insight, some fundamental issues still remain unsolved, such as, typically, the apparently unavoidable break down of Lorentz invariance as a result of the presence of nontrivial interactions. The expectation that ultraviolet (short distance) divergences would be  removed or  lessened, has been partly and in some case fully fulfilled, but generally,  the models investigated exhibit a strange mixing of ultraviolet and infrared divergences.  In the case when UV divergences disappear completely, the prize to pay for this  positive feature lies in serious difficulties in taking an adiabatic limit in time.

\subsection{Free fields and ``local algebras'' on QST}
In the approaches to QFT on Quantum Spacetime investigated by a number of the present authors, the free field equation remains unchanged, and therefore, the free massive bosonic quantum field on QST  can be understood as follows: after evaluation in a (suitable) state on QST, one obtains an operator on the ordinary Fock space $\mathcal H$ by the assignment
\begin{equation}\label{eq:QF}
\phi(\omega):=\frac{1}{(2\pi)^{3/2}}\,\int(\psi_\omega(k)\, a(\vec k)+\psi_\omega(-k)\, a(\vec k)^*)d\Omega^+_m(\vec k)=\varphi(\widehat{\psi_\omega})
\end{equation}
where $\psi_\omega(k)=\omega(e^{iq_\mu k^\mu})$ is the corresponding (inverse Fourier transformed) Wigner function, and where $\varphi$ is the quantum field on classical spacetime, $ \widehat{\phantom{\  }}$ denotes the Fourier transform, and  $d\Omega^+_m(\vec k)$ is the Lorentz-invariant measure on the positive mass shell as usual. For definiteness, the set of states might be chosen to be such that the resulting Wigner functions are Schwartz functions\footnote{This set is nonempty, as the Gauss function is the Wigner function of  the best localized states.}. Short-hand notation for the above construction is the formula~\eqref{eq:phiq} for $\phi(q)$ found in the Introduction.

One obtains a fully Poincar\'e covariant field, which gives rise to a Poincar\'e covariant  net of local algebras, as a map 
\[
E  \mapsto  \mathfrak A (E) \subset B(\mathcal H)
\]
which assigns to selfadjoint idempotents $E$ in the Borel completion of the C*-algebra $\cal E$ of Quantum Spacetime the von Neumann algebra generated by the (bounded functions of the appropriate self adjoint extensions of the real and imaginary parts of the) field operators in~\eqref{eq:QF}, when $\omega$ has support in $E$, i.e. $\omega (E) = 1$.

This map would be covariant: if ${\mathcal  {P}}$  is the covering group of the Poincar\' e group, and $\tau$, $\alpha$ respectively denote its action on the C*-algebra of Quantum Spacetime $\mathscr E$, extended by normality to the Borel completion,  and on the C*-algebra $\mathfrak F$ of field operators, then
\begin{equation}\label{eq:covariant}
\alpha _L  \mathfrak A (E)   =   \mathfrak A (\tau _L (E))              
\end{equation}
But Locality breaks down: if $\omega$ is translated by $a$ in a spacelike direction, even if $\omega$ is optimally localized,  the commutator between $\phi(\omega)$ and $\phi(\omega_a)$ is never zero. But, as explicitly computed for the typical case of free massless fields, it vanishes as a Gaussian of Planckian width as $a$ goes to spacelike infinity.  
%\[
%\phi(q):={1\over(2\pi)^{3/2}}\,\int(e^{iq_\mu k^\mu}\otimes a(\vec k)+e^{-iq_\mu k^\mu}
%\otimes a(\vec k)^*)d\Omega^+_m(\vec k)
%\]

Therefore the fields are no longer local -- which is perhaps to be expected  on QST --  but only at Planckian separations, for the free fields. It is not clear, however, if this results in a violation of causality at large scales in presence of interactions.  

Moreover, as we do not know how to deal with interactions in a Lorentz covariant way, we cannot be sure that a covariant net as in~\eqref{eq:covariant} can still be a picture of an interacting theory.

But even if it were, the formalism would still miss an essential ingredient to be significant: a clear cut algebraic property which replaces Locality and reduces to it in the limit where the Planck length is neglected. As Locality does in the classical Minkowski case~\cite{dopl17}, this axiom ought to imply most conceptual features of QFT on Quantum Spacetime, independently of the specific form of the interactions.

An indication of how radically new ideas are needed here is given by the dependence of the local algebras from our choices, already in the case of free fields.
 
If, as an example, we let $E(\lambda)$ denote the spectral family of  $q_0 ^2 + q_1 ^2 + q_2 ^2 + q_3 ^2$, whose spectrum is the half line with minimum $2$, and consider the local algebra  $ \mathfrak A (E(2)) $, there will be only one function $k   \mapsto  \omega (e^{i qk} )$ for all states $\omega$ on $\mathscr E$ such that $\omega (E(2)) = 1$. This function is a Gaussian. Hence, in the case of a single scalar and neutral field, the local algebra   $ \mathfrak A (E(2)) $ will be generated by a single self adjoint operator (with spectrum the real line, pure Lebesgue), and hence isomorphic to the commutative von Neumann algebra of  the Lebesgue $L^\infty$ complex functions on the unit circle (on the other hand, in the case of  finitely many generating  fields, all of Fermi type, it would be finite dimensional).

But if we generate a von Neumann algebra with the translates of that algebras over any tiny neighborhood of the origin  in the translation group, we find all bounded operators~\cite{dopl15} (see also~\cite{dopl16}).

\subsection{Perturbation theory}

When putting a quantum field theoretic model on Quantum Spacetime, several choices have to be made. In the absence of a good notion of locality, most publications have focused on perturbative approaches. Even so, the ordinary setup allows for a number of different generalizations. While on Minkowski space a number of approaches turn out to be equivalent (inductive construction of time-ordered products in the sense of Epstein and Glaser, Yang-Feldman approach, Dyson series, even Feynman graphs calculated via the Wick rotation), this ceases to be true on Quantum Spacetime. 

For one thing, only the Dyson series and the Yang-Feldman approach  seem to be even definable on Quantum Spacetime (where time does not commute with the space coodinates). And it then seems that, even on the simplest model of Quantum Spacetime, they yield theories which are inequivalent. Both approaches, however, share the feature that that they were based on the introduction of a commutative time parameter  $t$ -- in the Hamiltonian approach this was caused by taking a partial trace on the algebra,  
 \begin{equation}\label{eq:hamiltonian}
\mathcal H_I(t)=\int_{q^0=t}d^3 {\boldsymbol q}\  H(q)
\end{equation}
to define the interaction Hamiltonian $\mathcal H_I(t)$, and in the Yang-Feldman approach, such a time $t$ was introduced in order to define the incoming field and to even formulate the initial value problem. Here, the interacting field is calculated recursively, as a formal power series in the coupling constant, formally written (for the massive Klein-Gordon field), for the simplest choice of an interaction term,
\[
(\square + m^2)\phi = - g \phi^{n-1}, \quad
\phi =\sum_{k=0}^\infty g^k \phi_k.
\]
Here, the Klein-Gordon operator is defined via $\partial_\mu f(q):=\frac d {dt} f(q+t e_\mu I)|_{t=0}$. Fixing the intial condition by assuming that for $t \rightarrow -\infty$, the field $\phi(q+te_0 I)$ is the free field, the power series starts with the free field $\phi_0$, while higher orders are calculated as convolutions with the retarded propagator $G_{ret}$ of the ordinary Klein-Gordon equation, e.g.,
\[
\phi_1(q)=\int g(x) G_{ret}(x) \phi_0^{n-1}(q+xI) \,dx,
\]
with an infrared cutoff given by an $x$-dependent coupling constant $g$. Of course, the need for renormalization occurs here, since products of (even free) fields are ill-defined. Different methods of defining the interaction term have been investigated in e.g.~\cite{zahn:IRYF}, \cite{bdfp:quasi}. 

In the Hamiltonian formalism, on the other hand, it is important to note that in the expression for $\mathcal H_I(t)$ products of field operators appear which are spread in space and in time with a non local kernel, which is produced by the quantum nature of spacetime, see Section~\ref{subsec:interaction} below. Thus the time argument for the fields is not the parameter in $\mathcal H_I(t)$, but in the Dyson expansion for the $S$-matrix
\begin{equation}\label{eq:dyson}
\begin{split}
S &= T \exp\left[i\lambda \int_{-\infty}^{\infty}dt\, \mathcal H_I(t)\right]\\
& = I + \sum_{n=1}^{+\infty}\frac{(i\lambda)^n}{n!}\int_{-\infty}^{+\infty} dt_1\dots\int_{-\infty}^{+\infty} dt_n \,T\left[\mathcal H_I(t_1)\dots \mathcal H_I(t_n)\right]
\end{split}\end{equation}
the {\textit{time ordering $T$ has to be performed in terms of the arguments of the factors $\mathcal H_I(t)$}}, and not in terms of the time arguments in the fields~\cite{dopl1}. Otherwise an unjustified violation of unitarity is introduced~\cite{dopl4}. This prescription can be summarized in modifies Feynman rules~\cite{dopl3}.

\subsection{Interaction terms}
\label{subsec:interaction}
The next choice, which turns out to be just as delicate, is the generalization of even so simple an interaction term as $\phi^n$. We do not comment on gauge theories here, see however, e.g.~\cite{zahn:QED} and the comments on covariant coordinates and gauge invariant quantities in~\cite{dopl9}. 

The first possibility that comes to mind is to use the product in the Quantum Spacetime C*-algebra $\mathscr E$ to define $\phi(q)^n$. If this prescription is used, in the Dyson series approach, to define the Hamiltonian density $H(q)$ appearing in~\eqref{eq:hamiltonian}, it turns out that the resulting $\mathcal H_I(t)$ is still a function of the commutators $Q$.

In terms of interpretation, this means that besides the localization, an experimentalist would also have to specify which measure on the spectrum  $\Sigma$ of the centre he prepared. This problem would equally show up in the Yang-Feldman approach.

This cannot be solved by evaluating a Lorentz invariant state on the center, for the  Lorenz group is not amenable. Already in~\cite{dopl1} it was proposed  to use a distinguished state on the centre  in order to lessen this problem. More specifically, if $H(q, \sigma)$ denotes the evaluation of the Hamiltonian density at the point $\sigma \in \Sigma$, and $d\mu$ the rotation invariant regular probability measure on $\Sigma$ carried by the base $\Sigma _1$, in~\eqref{eq:dyson} the following expression was used
\begin{equation}
\mathcal H_I(t)=\int d\mu (\sigma )\int_{q^0=t}d^3 {\boldsymbol q}\  H(q, \sigma),
\end{equation}
but of course the \emph{ad hoc} choice of $d\mu$ breaks Lorentz invariance.

Note that, by power counting arguments,  the resulting $\phi^3$  theory was shown to be finite in this frame~\cite{dopl18}.

It must be mentioned that the twisting of the product of functions of $q$ caused by non commutativity suggested a very interesting approach initiated in~\cite{gandalf:twist, buchholz:Twist}. This framework of so-called warped products still holds potential to be an effective tool in the construction of  two dimensional models, with non trivial $S$ matrix; which can even be preassigned as a phase function in two particles elastic scattering, solving the inverse scattering problem in terms of wedge local algebras. In these approaches, locality is replaced by the weaker notion of wedge locality.

Coming back to QFT on Minkowski QST, in order to specify the quantum  Hamiltonian density $H(q)$ in~\eqref{eq:hamiltonian}, apart from using the product of $\mathscr E$, one sees that products of fields on QST, which generalize the ordinary interaction term, can now be defined in various ways, of which we mention two. 

The first one, originally adopted in the above mentioned works, relies on the interpretation that an interaction is produced by bringing fields close to each other -- in the end to bring them to coinciding points (at the cost, of course, of having to renormalize the corresponding term). This is the classical Wick procedure. But on QST it is not allowed to bring independent events at a coinciding point.
 
 Thus, in our framework, it is natural to redefine this limit of coinciding points using the quantum diagonal map introduced in Section~\ref{subsec:diagonal}
%subsection{Many events and the Diagonal map}
above.

A (classical) interaction term $\phi^n(x)$ is then replaced by 
\[
:\phi(\bar q)^n:_Q\; = E^{(n)}\big(:\phi(q_1)\phi(q_2)\dots\phi(q_n):\big),
\]
with $E^{(n)}$ as in equation~\eqref{eq:diagonal} %2.8 in version 7
and with the actual dependence on the quantum coordinate $\bar q$ of characteristic length $1/\sqrt{n}$ (the mean coordinate) already spelled out explicitly. 

The interaction Hamiltonian on the Quantum Spacetime is then given by
\[
\mathcal H_I(t)=\lambda\int_{q^0=t}d^3\bar{\boldsymbol{q}} :\phi(\bar q)^n:_Q
\]

This expression is independent of the commutators $Q^{\mu, \nu}$, hence no \emph{ad hoc} integration on $\Sigma$ is needed. But the definition of the quantum diagonal map chooses a particular Lorentz frame, hence Lorentz covariance is broken {\it ab initio}.

The above choice leads to a 
unique prescription for the interaction Hamiltonian on Quantum Spacetime. When used in the Dyson perturbative expansion for the $S$ matrix, this gives the same result as the  {\textbf{\textit{effective non local Hamiltonian}}} determined by the kernel
$$
\exp\bigg\{-\frac{1}{2}\sum_{j,\mu}{a_j^\mu}^2\bigg\}
\delta^{(4)}\bigg(\frac{1}{n}\sum_{j=1}^n{a_j}\bigg).
$$
The corresponding perturbative Gell-Mann and Low formula is then {\textbf{free of ultraviolet divergences}} at each 
term of the perturbation expansion~\cite{dopl5}. 

However those terms have a 
meaning only after a sort of adiabatic cutoff: the coupling constant should be changed to a 
function of time $\lambda_\tau$, rapidly vanishing at infinity, say depending upon a cutoff time $\tau$, i.e., the Gell-Mann and Low formula for the time-ordered products of the interacting field of the effective non local theory should read
$$
 \frac{ T\left(\phi(x_1)\dots\phi(x_n)\exp\left[ i\int_{-\infty}^{+\infty}dt\,\lambda _{\tau} (t) \mathcal H_I(t) \right]\right)}{\left\langle T\exp\left[i\int_{-\infty}^{+\infty}dt\,\lambda _{\tau} (t) \mathcal H_I(t) \right]\right\rangle_0}
$$
where the vacuum-vacuum contributions have to be divided out as usual, and where $T$ indicates the time ordering, of course with respect to the $t$-values in the expansion of the exponential, not the time values in the arguments of the field operators, as already remarked above.

Thus this prescription  leads to an ultraviolet finite theory, thereby finally fulfilling one of the original hopes of the whole approach. However, it remains to be shown that the adiabatic limit in time can be performed; otherwise, ultraviolet-infrared mixing problems cannot be excluded. This is an open problem, and there are indications that the limit might not exist.

Moreover, of course, a $\lambda_P$-dependent {\it finite renormalization}  would be needed anyhow, otherwise the results would not have any physical meaning, for they would include meaningless large contribution, divergent in the limit of classical Minkowski space.

From this perspective, a major open problem, anticipated in the Introduction,  is the following. 
Suppose we apply this construction to the renormalized Lagrangean of a theory which is  renormalizable on the ordinary Minkowski space, with the counterterms defined by that ordinary theory, and with finite renormalization constants depending upon both the Planck length $\lambda_P$ and the cutoff time $\tau$. 
Can 
we find a natural dependence such that in the limit $\lambda_P  \rightarrow 0$ and  $\tau \rightarrow \infty$ we get back the ordinary renormalized Gell-Mann and Low expansion on Minkowski space? 

This should depend upon a suitable way of performing a joint limit, which hopefully yields, for the physical value of  $\lambda_P$, to a result  which is essentially independent of $\tau$ within wide margins of variation, and can be taken as source of predictions to be compared with observations.

% Could be dropped. BTW was accepted for publication by  Europhysics B, but I never followed it up ... the editor who accepted it, is now retired... 

The other possibility for obtaining an interaction term is to consider auxiliary variables $x_i \in \mathbb R^4$ to define fields at separate points $q+x_iI$ in Quantum Spacetime, and to define the limit of coinciding points by letting $x_i\rightarrow x$, i.e., using this set of  commutative extra parameters. This was motivated mostly by the fact that, in the Yang-Feldman approach, such commutative separations occur anyhow.  
%(the quantum coordinate takes the role of a n extra degree of freedom). 
Also, it makes mathematically precise the idea that after evaluation  in a state on QST, one gets an ordinary operator valued  tempered distribution, similar to what one has in the Wightman formalism. In fact, after the choice of a localization state $\omega$, one considers as the $n$-fold tensor product of a quantum field on QST the tempered distribution
\[
\mathcal S(\mathbb R^{4n}) \ni g\mapsto \phi^{\otimes n}(\Psi_\omega\times g)
\]
where $\times$ is the convolution, and $\Psi_\omega(k_1,\dots,k_n)=\omega(e^{i(\sum_j k^\mu_j )q_\mu})$. Formally, this corresponds to considering products
\[
\phi(q+x_1)\cdots \phi(q+x_n)
\]
The crucial point is that one can now give a precise notion of what a local counterterm should be. The resulting Wick products which are defined by subtracting only such local counterterms were conjectured to be well-defined in the limit of coinciding points -- the proof which was sketched in~\cite{bdfp:quasi}  has been superseded by general considerations on twisted products of 
tempered distributions, which are currently being applied. But unfortunately this cannot be the end of the story. Some of the unsubtracted terms, even if finite for non zero values of the Planck length, are bound to diverge as that length is allowed to tend to zero. This means that they would contribute with possibly very large unphysical values, which ought to be removed by a finite renormalization.

Moreover, it turned out that this approach leads (in the Yang-Feldman approach)  to a strange dispersion relation (modified in the infrared), which cannot be absorbed by local counterterms. Furthermore, it was shown later~\cite{bahns:uvir}, that in the Hamiltonian formalism at least, the approach also exhibits a mixing of ultraviolet and infrared divergences\footnote{Note however, that in a Euclidean realm at least, there is hope that an infrared-cutoff model, the so-called  Grosse-Wulkenhaar model might have a chance to be resummable and thus give way even to a constructible theory.}.

\medskip
 At the heart of these problems seems to be the fact that we cannot control the effects of noncommutativity al large scales. In particular, we cannot control how would those effects cumulate at higher and higher orders of the perturbation expansion, and decide whether they would keep being sensible only at Plackian distances. To understand these issues better, seems to be one of the essential points to better understand quantum field theory in QST.

But, at a more fundamental level, the difficulties  with Lorentz covariance posed by the non triviality both of the center 
of the algebra of QST and of the action on it of the Lorentz group, might be a spy of the need of a more dynamical meaning of the commutators. As mentioned in the Introduction (cf.~also next Section), Physics suggest that those commutators  should depend on the fields, hence they should be acted upon by the Lorentz group in a more essential way. This might be the key to solve the problems with the correct definition of covariant interacting theories; however, in a scenario of which the only  thing which is clear is that it would be extremely difficult to treat. 

\section{Quantum Spacetime and Cosmology} 
\subsection{Beyond Minkowski: a dynamical Quantum Spacetime scenario}
\label{subsec:dynamicalqst}
The model of Quantum Minkowski Spacetime presented in the previous sections should be thought of as  a geometric background for Quantum Field Theory, which is more realistic than standard Minkowski Spacetime, as it implements in the noncommutative nature of the underlying geometry some of the limitations to localisability of events dictated by our present understanding of the basic principles of Quantum Mechanics and General Relativity. As shown above, the development of Quantum Field Theory on it allows us to avoid at least some of the problems and contradictions which we are otherwise bound to meet on commutative Minkowski.

As already mentioned, such a model seems to be sufficient for describing the typical regime of Particle Physics, in which the large scale spacetime structure is expected to have essentially no effect on particle collisions in an accelerator, even at very high energies. 

On the other hand, it is widely believed that a quantum description of Gravity becomes of relevance near classical gravitational singularities, e.g., at cosmological times smaller than the Planck time $t_P \simeq 10^{-43} s$, where it could provide a better understanding of the initial state of the universe. Moreover, it can be foreseen that gravitational effects that demand for such a quantum description can have observational consequences, for instance in the structure of the Cosmic Microwave Background. It seem therefore compelling to extend the analysis of the quantum structure of spacetime to the curved case, also in view of the fact that it is conceivable that Quantum Spacetime may serve as a more suitable background for Quantum Gravity too.

If we turn then to the consideration of a generally curved (commutative) spacetime and of quantum fields propagating on it, it is to be expected that the energy density of the prevailing quantum state affects the Spacetime Uncertainty Relations, as shown, e.g., by the argument presented in Section~\ref{sec:why}. Since this energy density determines the dynamics of spacetime itself, through Einstein's Equations, we are led to the conclusion that, on an arbitrary spacetime, the Spacetime Uncertainty Relations, and therefore the commutator between the coordinates of a generic event, should depend on the underlying metric tensor. This leads us to a scenario where the equations of motion of the system should then become (in natural units)~\cite{dopl7}:
\begin{align}
[q_\mu, q_\nu] &= i Q_{\mu\nu}(g), \label{eq:dqstcommutator}\\
R_{\mu\nu}-\frac{1}{2}g_{\mu\nu} R &= 8\pi T_{\mu\nu}(\phi), \label{eq:dqsteinstein}\\
F(\phi) &= 0, \label{eq:dqstfield}
\end{align}
where $\phi$ denotes the collection of the quantum fields under consideration, which should be thought as functions of the $q_\mu$'s, $T_{\mu\nu}(\phi)$ is their stress-energy tensor, and the last equation is symbolic for the fields' equations of motion (where the metric $g$ also appears via the covariant derivatives).

In order to turn this general picture into a model apt to perform actual calculations, it is of course necessary, among other things, to investigate more closely the possible form of the right hand side of~\eqref{eq:dqstcommutator}. This is of course a hard problem. Which maybe ought to be tackled without forgetting that  keeping the familiar form of~\eqref{eq:dqsteinstein} all the way down to the Planck scale is a terrific extrapolation: the experimental verification of Newton's law is not available for distances shorter than few millimetres.

Having said that, and lacking any clue on the possible modifications of Gravity at small scales, the simplest thing to do in order to study the possible form of the right hand side of~\eqref{eq:dqstcommutator} is to try to generalise the original derivation of the Spacetime Uncertainty Relations in~\cite{dopl1} to a generic curved spacetime treating the gravitational field in the semiclassical approximation. This means that one should estimate the backreaction of spacetime to the localisation of the state of a quantum field propagating on it, in order to detect the formation of trapped surfaces enclosing the localisation region. The first problem which arises is that the concept of energy, which enters the argument of~\cite{dopl1} through Heisenberg's Uncertainty Principle, is in general ill-defined on a curved background. Moreover it seems advisable to avoid also the other sharp simplifications made there, as, e.g., the use of the linearised form of the Einstein Equations to derive limitations which are relevant precisely in the extremely relativistic regime, where the linear approximation cannot be expected to be a good one. Or the use of a crude criterion, such as $g_{00} > 0$, for the non-formation of trapped surfaces.

\subsection{Localisation on a spherically symmetric spacetime}
\label{subsec:spherical}
The problems pointed out above have been solved in~\cite{dopl13} in the case of a spherically symmetric background and a spherically symmetric localisation region. The result, not surprisingly, is that in order to prevent the formation of trapped surfaces, the spatial sphere of localisation should have a radius  whose proper length is bounded below by a constant of the order of the Planck length. 

More specifically, consider a globally hyperbolic spacetime $M$ which is spherically symmetric. This means that $M$ is diffeomorphic to $I \times \mathbb{R}_+ \times \mathbb{S}^2$, with $I \subset \bR$ an open interval, and that the metric on $M$ takes the form
\begin{equation}\label{eq:sphericmetric}
ds^2 = -A(u,s)du^2 -2 ds\,du + r(u,s)^2 d\bS^2.
\end{equation}
The coordinates $(u,s) \in I\times \bR_+$ are the so called \emph{retarded coordinates}, and they have the following geometrical meaning. The coordinate $u$ is the proper time along the worldline $\gamma$ spanned by the centre of the spherical symmetry, while $s$ is the affine parameter along the future pointing null geodesics which emanate from the point $\gamma(u)$, normalised in such a way that the scalar product between the tangent vector to the considered geodesics and to $\gamma$ is one. The collection of all the lightlike geodesics emanating from $\gamma(u)$ forms a cone in $M$ which we will denote by $C_u$.

The surface of the spatial 2-sphere described by the points of $M$ at fixed $(u,s)  \in I\times \bR_+$ is given by $4\pi r(u,s)^2$, and, as intuitively clear, a trapped surface occurs when this quantity is decreasing with increasing $s$ at fixed $u$, as this means that the geodesics spanning $C_u$ are focusing. Thus, in order to detect the emergence of trapped surfaces, it is necessary to study the 
rate of change of this quantity. The latter is measured, along a fixed cone $C_u$, by the \emph{expansion parameter of null geodesics} $s \mapsto \theta(s)$, whose evolution is governed by the \emph{Raychaudhuri equation} (see, e.g.,~\cite{mor1}), which, under the present symmetry assumptions, reads
\begin{equation}\label{eq:raycha}
\dot \theta = -\frac{\theta^2}{2}- R_{ss}, \qquad \theta \sim \frac{2}{s}\quad\text{for }s\to 0^+.
\end{equation}
Here, $R_{ss}$ is the $s$-$s$ component of the Ricci tensor, which, due to spherical symmetry, is only dependent on $(u,s)$.

Consider now a scalar, massless, conformally coupled quantum field $\phi$ propagating on $M$, a background metric $g^{(0)}$ of the form~\eqref{eq:sphericmetric}, and an initial Hadamard state $\omega$ on the *-algebra $\mathcal{A}(M, g^{(0)})$ generated by the Wick monomials of $\phi$, in equilibrium with such background, namely a triple $(\phi, \omega, g^{(0)})$  satisfying the Klein-Gordon and the semiclassical Einstein Equations coupled together:
\begin{align}
\Box_{g^{(0)}} \phi &= 0, \label{eq:KG}\\
R^{(0)}_{\mu\nu}-\frac{1}{2}g^{(0)}_{\mu\nu}R^{(0)} &= 8\pi \omega(T_{\mu\nu}), \label{eq:semiclassEinstein}
\end{align}
being $T_{\mu\nu}$ the stress-energy tensor of the field $\phi$
(we refer the reader to~\cite{mor2} and references therein for a detailed discussion of a free scalar quantum field on a general globally hyperbolic background from the algebraic point of view). We emphasise that solutions to~\eqref{eq:semiclassEinstein} exist at least for spacetimes of cosmological interest~\cite{mor4, mor5}

The field $\phi$ is used to model an experiment of spherically symmetric localisation of an event on the background spacetime $(M, g^{(0)})$, and the state in which $\phi$ is prepared to perform such an experiment will be modeled by the following simple perturbation of $\omega$:
\begin{equation}
\omega_f(A) = \frac{\omega(\phi(f)A\phi(f))}{\omega(\phi(f)^2)}, \qquad A \in \mathcal{A}(M, g^{(0)}),
\end{equation}
with $f$ a spherically symmetric real smooth function whose support describes the localisation region of the event under consideration. Such a state is obviously not strictly localised in $\mathrm{supp} f$, and this entails that the limitations obtained on the size of the localisation region will be weaker that those deriving from a strictly localised one, whose energy density, at fixed total energy, will be larger.

These limitations arise in principle by considering the backreaction of the underlying metric to the localisation, i.e., the solution $g_{\mu\nu}$ to the semiclassical Einstein Equations with source the stress-energy tensor of $\phi$ (on the fixed background $g^{(0)}$) in the perturbed state $\omega_f$,
\begin{equation}
R_{\mu\nu}- \frac{1}{2}g_{\mu\nu}R = 8\pi\omega_f(T_{\mu\nu}),
\end{equation}
and imposing that, in accordance to the Principle of  Gravitational Stability, no trapped surface appears preventing signals from $\mathrm{supp} f$ to reach a distant observer. 

In practice, this is accomplished in~\cite{dopl13} by first evaluating the change in the expectation value of the stress-energy tensor $\langle T_{ss} \rangle_{f,0} := \omega_f(T_{ss})-\omega(T_{ss})$; then by fixing a cone $C_0$ containing $\mathrm{supp} f$ in its causal future and considering ~\eqref{eq:raycha} on it, where, in the right hand side,
\[
R_{ss} = 8\pi\omega_f(T_{ss}) = R_{ss}^{(0)} + \langle T_{ss} \rangle_{f,0}
\]
(remember that $g_{ss} =0$); and finally by requiring that its solution remains positive for all $s > 0$. This, according to the above discussion, entails that no trapped surface appears in the future of $C_0$. We notice explicitly that in this procedure no use of ill-defined concepts like energy is made, as the estimate of $\langle T_{ss} \rangle_{f,0}$ is solely a consequence of the free field properties, namely of the CCR.

The outcome of this discussion is summarised in the following theorem.
\begin{theorem}[\cite{dopl13}]
Under the above hypotheses and notations, assume moreover that:
\begin{itemize}
\item[(i)] $R_{ss}^{(0)} \geq 0$  on $C_0$;
\item[(ii)] there exists a constant $C > 0$ such that %for all smooth functions $f$ with compact support in the future of $C_0$,
\[
|\omega(\phi(f)\phi(f))| \leq C\|s \psi_f\|_{L^2(C_0)}\|\partial_s(s\psi_f)\|_{L^2(C_0)},
\]
where $\psi_f := \Delta(f)|_{C_0}$ is the restriction to $C_0$ of the image of $f$ under the causal propagator $\Delta$ of equation~\eqref{eq:KG};
\item[(iii)] defining $s_2 > 0$ to be the value of the affine parameter such that the points of $C_0$ in the past causal shadow of $\mathrm{supp} f$ satisfy $s < s_2$, there exists an $s_1 < s_2 < \frac{3}{2}s_1$ such that
\[
\|\partial_s \psi_f\|^2_{L^2(C_0)} \leq 8\pi \int_{s_1}^{s_2} |\partial_s\psi_f|^2\,ds.
\]
\end{itemize}
Then, for the expansion parameter $\theta$ to be positive on $C_0$, it is necessary that $s_2 \geq  \bar s:=1/\sqrt{12 C}$.
\end{theorem}

\begin{figure}
     \begin{picture}(50,80)(-100,0) %(54,90)(0,0)
       \put(20,0){\includegraphics[width=5cm]{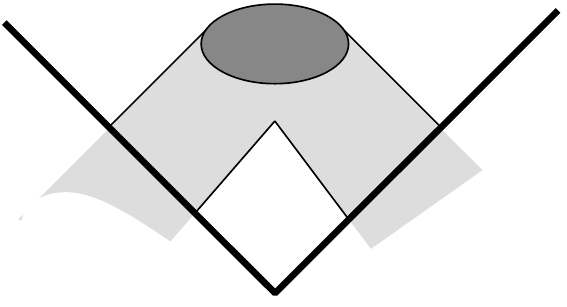}}
       \put(160,58){\small $C_0$}
       \put(80,85){\small $\mathrm{supp} f$}
       \put(105,10){\small $s_1$}
       \put(137,42){\small $s_2$}
       %\put(-10,40){\tiny $supp\, \Delta(f)\cap J^{-}(O)$}
     \end{picture}
     \caption{Past null shadow of $\mathrm{supp} f$.}
     \label{fig:shadow}
\end{figure}

We remark that assumption (i) is verified at least in all reasonable cosmological spacetimes, and that assumption (ii) is satisfied (with $C=1$) by the massless Minkowski vacuum~\cite[Appendix]{dopl13}, and that a similar property holds for a large class of Hadamard states on curved backgrounds~\cite{mor3}. Finally, assumption (iii) appears to be reasonable if $s_1, s_2$ are related to the past null shadow of $\mathrm{supp} f$ like in Fig.~\ref{fig:shadow},
due to the fact that the dominant contribution to $\|\partial_s \psi_f\|^2_{L^2(C_0)}$ comes from the singularities of the causal propagator $\Delta(x,y)$ for lightlike separations of the arguments. 

Thus, for the localisation experiment to be physically realisable, the size of the localisation sphere, as measured in terms of the affine parameter, has to be bounded below by some constant $\bar s$ of order 1. We obtain in this way a generalisation to a curved spherically symmetric space-time of the particular case of the Spacetime Uncertainty Relations in which all the uncertainties are of the same order of magnitude. In order to get a full set of Spacetime Uncertainty Relations, it would be of course necessary to treat the case in which $\mathrm{supp} f$ is not spherically symmetric.

 The achieved result, anyway, means in  particular that in a flat Friedmann-Robertson-Walker (FRW) background (which is spherically symmetric with respect to every point), with metric, in spatial spherical coordinates, $ds^2 = -dt^2+a(t)^2[dr^2+r^2d\mathbb{S}^2]$, the size of a localisation region centred around an event at cosmological time $t$, measured by the radial coordinate $r$, must be at least of order $1/a(t)$ (and therefore of order 1 in terms of proper length). Thus this gives further support to the expectation that the Spacetime Uncertainty Relations are affected by the background metric, and therefore to~\eqref{eq:dqstcommutator}.

\subsection{Backreaction on Quantum Spacetime and the horizon problem}
\label{subsec:horizon}
The above derived behaviour of the effective Planck length on a flat FRW spacetime suggests of course that the acausal effects induced by the quantum structure of spacetime should  become more important near the Big Bang, when $a(t) \to 0$, in agreement with previous remarks. In particular, as seen also in Sec.~\ref{sec:QFT}, it can be expected that this results in a high-energy modification of the product of quantum fields at the same spacetime point, and in particular of the stress energy-tensor. In the spirit of the scenario outlined in Sec.~\ref{subsec:dynamicalqst}, this should entail, in turn, a modified cosmological evolution.
 
In~\cite{dopl13} this issue has been analysed in more detail in the simplified situation of a universe only filled with radiation (modeled by a massless scalar field). In order to circumvent the problem of not knowing the explicit form of the commutation relations~\eqref{eq:dqstcommutator} implementing the Spacetime Uncertainty Relations on a generic background, and therefore without a full-fledged QFT on the resulting Quantum Spacetime algebra, the following strategy was adopted: first evaluate the modification of the stress-energy tensor on ordinary Minkowski Quantum Spacetime, then use the conformal isometry of (commutative) flat FRW with Minkowski to propose an \emph{ansatz} for the stress-energy tensor on (the yet unknown) Quantum FRW Spacetime, and finally solve the semiclassical Einstein Equation with source given by the expectation value of such modified stress-energy tensor in a thermal state, used as a simple approximation of the initial hot state of the universe, whose relics we see today as the Cosmic Microwave Background (CMB). The interesting result of this analysis is that, although the Big Bang singularity is still present in this model, the scaling behaviour of radiation density near the singularity in significantly modified, in a way such that the resulting cosmological evolution avoids the horizon problem of standard cosmology. 
 
More in detail, given a free massless scalar field $\phi$ on Minkowski Quantum Spacetime $\mathscr{E}$, its energy density is defined by replacing the coinciding point limit with the quantum diagonal map of Section~\ref{subsec:diagonal}:
\[
:\rho:_Q(\bar q) := E^{(2)}\left(:\partial_0\phi(q_1)\partial_0\phi(q_2):-\frac{1}{2}\eta_{\mu\nu}:\partial^\mu\phi(q_1)\partial^{\nu}\phi(q_2):\right).
\]
Consequently, the expectation value of $:\rho:_Q$ in the unique KMS state $\omega_\beta$ at inverse temperature $\beta > 0$ is easily calculated to be, in generic units,
\begin{equation}\label{eq:densityminkowski}
\omega_\beta(:\rho:_Q(\bar q)) = \frac{1}{2\pi^2}\int_0^{+\infty}dk\,k^3 \frac{e^{-\lambda_P^2k^2}}{1-e^{\beta k}},
\end{equation}
where the $\bar q$ dependence disappears in the right hand side because of translation invariance of $\omega_\beta$. The result differs from the analogous quantity on commutative Minkowski spacetime by the Gaussian damping at high energies in the integrand. 
 
Consider now a free, massless, conformally coupled field $\phi$ on flat FRW spacetime $M$. Introducing the conformal time $\tau = \int_{t_0}^t \frac{dt'}{a(t')}$, the metric of $M$ becomes $ds^2 = a(\tau)^2[-d\tau^2+d\boldsymbol{x}^2]$ and therefore $M$ is conformally isometric to a subset of Minkowski spacetime. This entails that the state $\omega_\beta$ induces a state $\omega^M_\beta$ on (the algebra generated by the Wick powers of $\phi$ on) $M$, by simply replacing, in its two-point function, $\beta$ with $\beta a(t)$. Accordingly, its physical interpretation (e.g., in the framework of~\cite{mor6}) can be seen to be that of a state describing local thermal equilibrium at inverse temperature $\beta(t) = \beta a(t)$.

This fact, together with the observation, made at the end of Sec.~\ref{subsec:spherical}, that the effective Planck \emph{proper} length is constant in time, i.e., it does not scale with $a(t)$, is at the basis of the following \emph{ansatz}~\cite{dopl13}: in passing from Quantum Minkowski Spacetime to Quantum Spacetime modeled on flat FRW, the only effective change on the expectation value of the energy density of $\phi$ is given by replacing $\beta$ in~\eqref{eq:densityminkowski} with $\beta(t) = \beta a(t)$.

The resulting expression for the energy density is therefore
\begin{equation}\label{eq:densityFRW}
\rho_\beta(t) := \omega_t \otimes \omega^M_\beta(:\rho:_Q(\bar q)) = \frac{1}{2\pi^2}\int_0^{+\infty}dk\,k^3 \frac{e^{-\lambda_P^2k^2}}{1-e^{\beta a(t) k}},
\end{equation}
where $\omega_t$ is a  state in which the cosmological time coordinate of $M$ is sharply localised, and the dependence on the other components of $\bar q$ disappears again due to the spatial translation invariance of $\omega^M_\beta$. It is easy to see that, while this expression is a negligibly small correction of the standard one on commutative flat FRW for $\frac{\lambda_P}{\beta a(t)} \to 0$, its asymptotic behaviour for $\frac{\lambda_P}{\beta a(t)}  \to +\infty$ is given by
\[
\rho_\beta(t) \sim \frac{C}{\beta a(t) \lambda_P^3},
\]
significantly different from the standard one $\sim 1/a^4$.

Thanks to the assumed symmetry of the metric, the semiclassical Einstein Equations reduce to the first Friedmann equation, which, for $a(t) \to 0$ (i.e., near the Big Bang), takes then the simple form
\[
\left(\frac{\dot a}{a}\right)^2 = \frac{c}{a},
\]
and has therefore solutions, in terms of the conformal time $\tau$, of the form $a(\tau) = (\alpha \tau + \beta)^{-2}$, from which one sees that the Big Bang occurs for conformal time $\tau \to -\infty$. This means that the singularity is the lightlike past boundary of the conformally related Minkowski spacetime, and thus in this spacetime every couple of points have been in causal contact at some time in the past after the Big Bang. This is to be compared with the standard cosmological evolution driven by a radiation field, where the Big Bang corresponds to some spacelike surface at finite conformal time $\tau = \tau_0$, which produces the horizon problem, illustrated by Fig.~\ref{fig:horizon}:
\begin{figure}
\centering
\includegraphics[width=5cm]{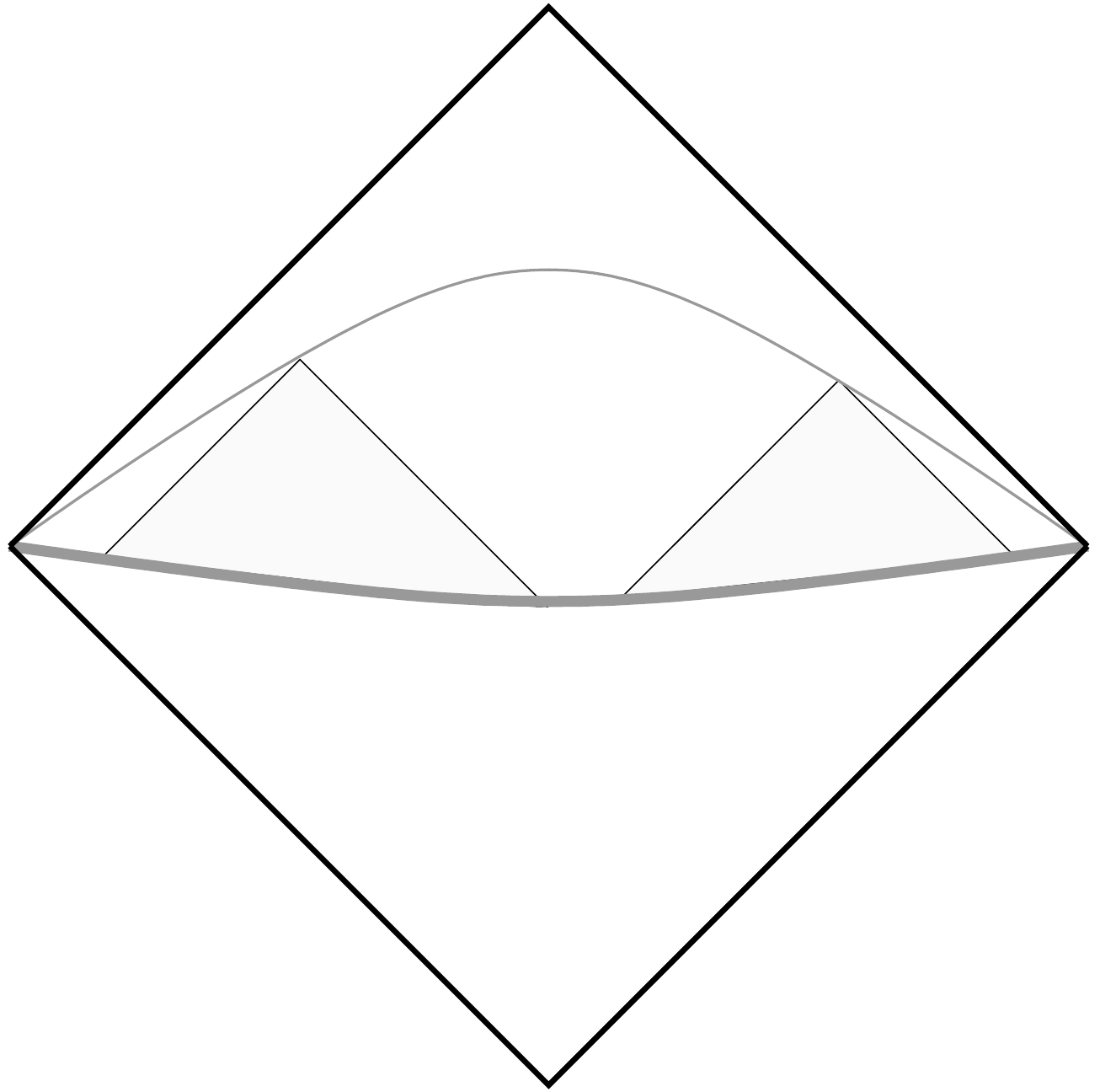}
\caption{The horizon problem.}
\label{fig:horizon}
\end{figure}
on any spacelike surface, there exists events which were never in causal contact since the Big Bang, which conflicts with the high degree of homogeneity of the CMB over the entire sky.

We recall that the commonly accepted solution to this problem, the inflationary scenario (see, e.g.,~\cite{mor7}), typically postulates the existence of an \emph{ad hoc} field, the inflaton, with a specific interaction, which has the role of driving a cosmological evolution without the horizon problem, and then decouples. In the model presented in~\cite{dopl13}, on the contrary, the field $\phi$ is just a free field, and the inflationary expansion is produced by the high-energy modification of its energy density caused by the quantum structure of spacetime, so that it can be expected that this is a generic feature, occurring also in more realistic situations of Standard Model fields interacting with a background ``Quantum FRW Spacetime''.

Finally, we stress that these results give further support to the discussion in Sec.~\ref{sec:why} motivating the scenario~\eqref{eq:dqstcommutator}-\eqref{eq:dqstfield}, and also agree with the heuristic argument~\cite{dopl7} which suggests to modify the Planck length on a curved background by, as a rough approximation, the factor $g_{00}^{-1/2}$. Such a rough argument points too to an infinite extension of non local effects near a singularity, so that, near the ÔÔBig BangÕÕ, thermal equilibrium would have been established globally.

\subsection{Further possible cosmological applications}
The findings reported above, although obtained in a semiclassical, oversimplified version of the scenario presented in Sec.~\ref{subsec:dynamicalqst} resulting from extrapolations of properties of QFT on commutative curved spacetime, provide a strong motivation for its further analysis.

Among the issues to be considered in a more refined framework, a prominent position is certainly taken by the analysis of the possible solutions given to the other basic problems of standard cosmology (as, e.g., the flatness problem), usually solved by inflationary models. Moreover, a study of the structure of CMB anisotropies induced by QFT on Quantum Spacetime promises to be a very stringent test, in view of the many experimental data which have become available in recent times~\cite{mor8}, which already put rather severe constraints on inflation~\cite{mor9}.

It is also worth mentioning that the apparent increase of the range of acausal effects near classical curvature singularities, together with the characteristic property of QFT on Quantum Spacetime of mixing short and long range effects, could also be expected to be at the root of the emergence of a non-zero Cosmological Constant, as a form of effective repulsion at very short distances~\cite{dopl7}. In another direction, the same feature points to a minimal size for black holes, where Hawking evaporation would stop. The minimal black holes would be stable, and fill the universe with a gas which would contribute to the dark matter. Due to the spherical symmetry involved, some indications on this issue could come from an adaptation of the arguments presented in Sec.~\ref{subsec:spherical}. The problem, however, would be displaced from the nature of dark matter to that of the possible formation of such black holes.

In an equally speculative attitude, it is conceivable that a dynamical Quantum Spacetime could solve the problem of maintaining Lorentz covariance in interacting QFT on it, and also that it could serve as a suitable geometrical background for the formulation of a consistent Quantum Gravity theory.

To conclude this section, we mention that a full set of Spacetime Uncertainty Relations for a flat FRW background has been obtained in~\cite{dopl12}, starting from an \emph{ansatz} on the formation of trapped surfaces which generalises exact results for spherically symmetric or equipotential surfaces, but still using Heisenberg Principle to evaluate the energy content of the localised quantum state. Moreover, an implementation of these Spacetime Uncertainty Relations is proposed in terms of concrete Hilbert space operators satisfying specific commutation relations. This could then be taken as a starting point for the formulation of a symmetry-reduced, semiclassical version of~\eqref{eq:dqstcommutator}-\eqref{eq:dqstfield}, in which some of the above mentioned problems could be addressed.

%\section{Nets of Local Algebras on Quantum Spacetime,  fates of Lorentz Covariance and Locality}


\begin{thebibliography}{99}

\bibitem{dopl1}
S.~Doplicher, K.~Fredenhagen, J.~E.ÜRoberts: ``The Quantum
Structure of Spacetime at the Planck Scale and Quantum Fields'', Commun.
Math. Phys. 172, 187-220 (1995);

\bibitem{dopl2}
S.~Doplicher, K.~Fredenhagen, J.~E.~Roberts:  ``Spacetime
Quantization Induced by Classical Gravity'', Phys. Letters B 331, 39-44
(1994);

\bibitem{dopl7}
S.~Doplicher: ``Space-time and fields: A Quantum texture'',   in Karpacz 2001, ``New 
developments in fundamental interaction theories'', 204-213, arXiv:hep-th/0105251;

\bibitem{dopl13}
S.~Doplicher, G.~Morsella, N.~Pinamonti: ``On Quantum Spacetime and the horizon problem'', J. Geom. Phys. 74 (2013), 196-210;

\bibitem{dopl14}
M.~P.~Bronstein: ``Quantum Theory of Weak Gravitational Fields'', (Republication from  Physikalische Zeitschrift der Sowjetunion, Band 9, Heft 2Ð3, pp. 140Ð157 (1936), English translation by M.A. Kurkov, edited by S. Deser) Gen. Relativ. Gravit. 44, 267Ð283 (2012), DOI 10.1007/s10714-011-1285-4;

\bibitem{ACV} D.~Amati, M.~Ciafaloni, G.~Veneziano, `Higher-order gravitational deflection and soft bremsstrahlung in planckian energy superstring collisions'',
Nucl.~Phys.~B 347, 550 (1990);

\bibitem{dopl11}
L.~Tomassini, S.~Viaggiu: ``Physically motivated uncertainty relations at the Planck length for an emergent non commutative spacetime''  Class. Quantum Grav. 28, 075001 (2011);

\bibitem{dopl12}
L.~Tomassini, S.~Viaggiu: ``Building non commutative spacetimes at the Planck length for Friedmann flat cosmologies'', Class. Quantum Grav. 31, 185001 (2014),  arXiv:1308.2767;

\bibitem{dopl5}
D. Bahns, S. Doplicher, K. Fredenhagen, G. Piacitelli:
``Ultraviolet finite quantum field theory on quantum space-time''
Commun. Math. Phys. 237, 221-241 (2003);

\bibitem{dopl9}
D. Bahns, S. Doplicher, K. Fredenhagen, G. Piacitelli:  ``Quantum Geometry on Quantum Spacetime: Distance, Area and Volume Operators'', Commun. Math. Phys  308, 567-589 (2011);

\bibitem{dopl6}
G. Piacitelli: ``Twisted Covariance as a Non Invariant Restriction of the Fully 
Covariant DFR Model'', arXiv:0902.0575 [hep-th];

\bibitem{dopl4}
D. Bahns, S. Doplicher, K. Fredenhagen, G. Piacitelli: 
``On the Unitarity problem in space-time noncommutative theories''
Phys. Lett. B 533, 178-181 (2002); 

\bibitem{dopl3}
G.~Piacitelli: ``Nonlocal theories: New rules for old diagrams'',
JHEP 0408: 031 (2004), arXiv: hep-th/0403055;

\bibitem{dopl15}
S.  Doplicher, K. Fredenhagen, unpublished;

\bibitem{dopl16}
C.~Perini, G.~N.~Tornetta: ``A scale covariant Quantum Spacetime'',  Rev. Math. Phys., to appear, arXiv:1211.7050;

\bibitem{dopl17}
S.~Doplicher: ``The principle of locality: Effectiveness, fate,
and challenges'',  J. Math. Phys. 51, 015218  (2010);

\bibitem{AC} G.~Amelino Camelia: ``Testable scenario for relativity with minimum length'', Phys.~Letters~B, 510, 255-263 (2001);

\bibitem{zahn:IRYF}C.~Doescher, J.~Zahn: ``Infrared cutoffs and the adiabatic limit in noncommutative spacetime'', Phys.Rev. D 73, 045024 (2006); 

\bibitem{bdfp:quasi}
D.~Bahns, S.~Doplicher, K.~Fredenhagen, G.~Piacitelli: ``Field theory on noncommutative spacetimes: Quasiplanar Wick products'', Phys. Rev. D 71, 1-12 (2005); 

\bibitem{zahn:QED}
J.~Zahn: ``Noncommutative (supersymmetric) electrodynamics in the Yang-Feldman formalism'', Phys. Rev. D 82, 105033 (2010), arxiv:1008.2309;

\bibitem{dopl18}
D.~Bahns: ``Ultraviolet Finiteness of the averaged Hamiltonian on the noncommutative Minkowski space'',  arXiv:hep-th/0405224;

\bibitem{gandalf:twist} 
H.~Grosse, G.~Lechner:  
``Wedge-Local Quantum Fields and Noncommutative Minkowski Space'', JHEP 0711, 012 (2007);

\bibitem{buchholz:Twist} 
D.~Buchholz, G.~Lechner, S.~Summers: ``Warped Convolutions, Rieffel Deformations and the Construction of Quantum Field Theories'', Commun. Math. Phys. 304, 95-123 (2011); 


\bibitem{bahns:uvir}
D.~Bahns: ``On the ultraviolet-infrared mixing problem on the noncommutative Minkowski space'', Ann. H. Poincar/'e, to appear;

\bibitem{mor1}
R.~M.~Wald: \emph{General Relativity}, Chicago University Press, 1984;

\bibitem{mor2}
R.~Brunetti, K.~Fredenhagen, R.~Verch: ``The generally covariant locality principle: A new paradigm for local quantum physics'', Commun. Math. Phys. 237, 31-68, (2003);

\bibitem{mor4}
N.~Pinamonti: ``On the initial conditions and solutions of the semiclassical Einstein equations in a cosmological scenario'', Commun. Math. Phys. 305, 563-604 (2011);

\bibitem{mor5}
N.~Pinamonti, D.~Siemssen: ``Global existence of solutions of the semiclassical Einstein equation for cosmological spacetimes'', Commun. Math. Phys., to appear, arXiv:1309.6303;

\bibitem{mor3}
C.~Dappiaggi, N.~Pinamonti, M.~Porrmann: ``Local causal structures, Hadamard states and the principle of local covariance in quantum field theory'',
Commun. Math. Phys. 304, 459-498 (2011);

\bibitem{mor6}
D.~Buchholz, I.~Ojima, H.~Roos: ``Thermodynamic properties of non-equilibrium states in quantum field theory'', Ann. Phys. 297, 219-242 (2002);

\bibitem{mor7}
P.~J.~E.~Peebles, \emph{Principles of physical cosmology}, Princeton University Press, 1993;

\bibitem{mor8}
Planck Collaboration: ``Planck 2013 results. I. Overview of products and scientific results'', A\&A, to appear, arXiv:1303.5602;

\bibitem{mor9}
A.~Ijjas, P.~J.~Steinhardt, A.~Loeb: ``Inflationary paradigm in trouble after Planck2013'', Physics Letters B 723, 261-266 (2013), arXiv:1304.2785.


%\bibitem{dopl8}
%S.~Doplicher: ``Quantum Field Theory on Quantum Spacetime'', 
%J. Phys. Conf. Ser. 53, 793-798 (2006), arXiv: hep-th/0608124;

%\bibitem{dopl10}
%S.~Doplicher: ``The Measurement Process in Local Quantum Theory
%and the EPR Paradox'', arXiv:0908.0480.


\end{thebibliography}
\end{document}